\documentclass{ptephy}

\usepackage{boites}
\usepackage{lineno}
\usepackage[normalem]{ulem}
\usepackage{color}

\def\lsim{\raise0.3ex\hbox{$\;<$\kern-0.75em\raise-1.1ex
\hbox{$\sim\;$}}}
\def\gsim{\raise0.3ex\hbox{$\;>$\kern-0.75em\raise-1.1ex
\hbox{$\sim\;$}}}

\begin{document}

\title{Constraining Majorana CP phase in precision era of cosmology 
and  double beta decay experiment
}

\author{\name{\fname{Hisakazu} \surname{Minakata}}{1,2,\ast}, 
\name{\fname{Hiroshi} \surname{Nunokawa}}{1,\ast}, 
and \name{\fname{Alexander} \surname{A. Quiroga}}{1,\ast}
}
\address{
\affil{1}{
Departamento de F\'{\i}sica, Pontif{\'\i}cia Universidade Cat{\'o}lica 
do Rio de Janeiro,  C. P. 38071, 22452-970, Rio de Janeiro, Brazil}
\affil{2}{Instituto de F\'{\i}sica, Universidade de S\~ao Paulo,
C.\ P.\ 66.318, 05315-970 S\~ao Paulo, Brazil} 
%
\email{hisakazu.minakata@gmail.com, nunokawa@puc-rio.br,
alarquis@fis.puc-rio.br} 
} 
\begin{abstract}
We show that precision measurement of 
(1) sum of neutrino masses by cosmological observation and 
(2) lifetime of neutrinoless double beta decay in ton-scale experiments, 
with supplementary use of 
(3) effective mass measured in single beta decay experiment, 
would allow us to obtain information on the Majorana phase of neutrinos.
To quantify the sensitivity to the phase we use the CP exclusion fraction, 
a fraction of the CP phase parameter space that can be excluded for 
a given set of assumed input parameters, a global measure for CP violation.
We illustrate the sensitivity under varying assumptions, from modest to optimistic ones, on experimental errors and theoretical uncertainty of nuclear matrix elements. Assuming that the latter can be reduced to a factor of $\simeq 1.5$ we find that one of the two Majorana phases (denoted as $\alpha_{21}$) can be constrained by excluding $\simeq 10-40\%$ of the phase space at $2\sigma$ CL  even with the modest choice of experimental error for the lowest neutrino mass of 0.1 eV. The characteristic features of the sensitivity to $\alpha_{21}$, such as dependences on the true values of $\alpha_{21}$, are addressed. 
\end{abstract}

\subjectindex{B52, B54, C04, C43, F23}

\maketitle


\section{Introduction}
\label{sec:intro}

After all the angles in the lepton mixing matrix \cite{MNS} and 
the mass squared differences of neutrinos are determined by 
the atmospheric, the solar, the reactor, as well as the accelerator 
neutrino oscillation experiments, we have entered into 
a new phase of neutrino 
physics~\cite{Kajita:2012vc,McDonald:2004dd,Inoue:2004wv}. 
Of course, we are still with the important unknowns in the lepton
sector, CP violating phase of the Kobayashi-Maskawa type~\cite{KM} and 
the neutrino mass ordering. Yet, we do have relatively clearer view 
of how these questions are to be settled, either through the ongoing 
searches, or by new measurement to be carried out by the powerful 
apparatus described in various experimental proposals to date.

It is quite possible that even after these unknown quantities are
measured by experiments the questions of somewhat different category 
will still remain: What are the absolute masses of neutrinos? 
Are neutrinos Dirac or Majorana particles? If Majorana, 
which values of the Majorana CP 
phases \cite{Schechter:1980gr,Bilenky:1980cx,Doi:1980yb} are 
chosen by nature? 
The first one, knowing the absolute mass scale, is crucial 
to understand physics behind the origin of neutrino mass, 
and the latter two are likely the key to understand 
baryon number asymmetry in the universe \cite{leptogenesis}.

In this paper, we focus on detectability of the Majorana CP phase in the light of informations to be obtained by cosmological observation and neutrino experiments in the coming precision era. In particular, in the near future the double beta decay experiments aim to cover entire region of parameters allowed by the inverted mass ordering,\footnote{
Throughout the paper the ``ordering'' of masses refers to that of the ones relevant for atmospheric neutrino oscillation, namely, normal (inverted) mass ordering corresponds to $m_2 < m_3$ ($m_2 > m_3$). 
} 
see Fig.~\ref{fig:m0nbb-m0}. We show that given the future perspective, when combined with precision measurement in cosmology, one can indeed expect some sensitivities, though limited ones, to one of the phases (denoted as $\alpha_{21}$, see section~\ref{subsec:double-beta}). We hope that this result can trigger renewed interests in the problem of detectability of the Majorana phase despite the dominant pessimism which prevailed in the past decades. See, 
e.g., Refs.~\cite{Minakata:1996vs,Bilenky:2001rz,Czakon:2001uh,Pascoli:2001by,Barger:2002vy,Nunokawa:2002iv,Pascoli:2002qm,Deppisch:2004kn,Joniec:2004mx,Pascoli:2005zb,Choubey:2005rq,Simkovic:2012hq} for the foregoing efforts which examined the sensitivity to the Majorana phase. 

We feel that the present moment is the particularly right time to revisit these issues because of the following reasons. Firstly, cosmological observations entered into the precision era so that their data became sensitive to the absolute masses of neutrinos better than the current bounds from laboratory experiments, as seen most notably by the results obtained by the Planck collaboration~\cite{Ade:2013zuv}. See also Ref.~\cite{Abazajian:2011dt} for overview and many relevant references. It makes us possible to employ a new strategy that the absolute neutrino mass scale can be constrained mainly by cosmology and we can use data of neutrinoless double beta decay experiments, with supplementary use of single beta decay data, to constrain the Majorana phases of neutrinos. Even more interestingly a discrepancy between the cosmic microwave background (CMB) and lensing observations about cluster correlations would prefer massive neutrinos with mass of a few tenth of eV \cite{Battye:2013xqa}. If confirmed, it would make the basis of our analysis more robust.

Secondly, as mentioned above, several new neutrinoless double beta decay experiments are in line to measure or constrain, the ``ee'' element of neutrino mass matrix in the flavor basis, so called the effective Majorana mass, to a few tens of meV~\cite{Albert:2014awa,Auger:2012ar,Gando:2012zm,Agostini:2013mzu,Arnaboldi:2002du,Hartnell:2012qd,Gaitskell:2003zr,Arnold:2010tu,Alvarez:2011my}. For reviews and description of the other projects, see e.g., Refs.~\cite{GomezCadenas:2011it,Giuliani:2012zu,Cremonesi:2013bma}. Needless to say, they are the prime candidates among various experiments that can measure or constrain the Majorana phases. We should also note that intensive efforts have been devoted in order to improve the calculations of the nuclear matrix element (NME) which is crucial to determine the value of effective Majorana mass from the measured half life time in double beta decay experiments~\cite{Rodin:2003eb,Rodin:2007fz,Caurier:1996zz,Caurier:2007wq,Simkovic:2009pp,Kortelainen:2007rh,Kortelainen:2007mn,Barea:2009zza,Barea:2012zz,Barea:2013bz,Menendez:2008jp}. See, also Ref.~\cite{Cremonesi:2013bma} for review of the recent progress in the calculations of NME. 

Thirdly, the uncertainties in the relevant lepton mixing angles get decreased dramatically recently. The error of $\sin^2 \theta_{12}$ is now only $\sim 4\%$\cite{Gando:2013nba}, and moreover it will be decreased to a sub-percent level by the future medium-baseline reactor experiments~\cite{Kettell:2013eos,Seo:2013swa}. 
The mixing angle $\theta_{13}$ whose value was unknown until recently 
before intriguing indication from T2K~\cite{T2K} by $\nu_{e}$ appearance
events, is now precisely measured \cite{An:2012bu,RENO}. 
The error of $\sin^2 \theta_{13}$ will soon reach to a even greater
precision, $\sim 5\%$ level. 
See Fig.~\ref{fig:m0nbb-m0} which shows how small is the effect of the
current uncertainties of the mixing parameters onto the allowed regions 
by double beta decay experiments. 
These developments are important to tighten up the constraint on 
Majorana CP phases and to place it on a firmer ground. 

In addition, the single beta decay experiment KATRIN \cite{Wolf:2008hf} (see also \cite{KATRIN}) will constrain neutrino masses to a level of 0.2 eV at 90 \% CL, or observe the effects consistent with masses of this order, in a manner quite complementary to cosmological observations. While enjoying the above new inputs, our analysis is nothing but a continuation of the numerous similar analyses attempted by the many authors, from which we are certainly benefited. 

To display the sensitivities to the Majorana CP phases in a global way,
we use in this paper the CP exclusion fraction $f_{\rm CPX}$
\cite{Machado:2013kya}, which is defined as a fraction of the CP phase
space that can be excluded at a certain CL for a given set of 
input parameters. 
We believe that it is a useful tool particularly in the initial 
era of search for effects of the Majorana CP phase, 
in which even the partial exclusion of allowed range of 
the phase would be very valuable.\footnote{
On the other hand, $f_{\rm CPX}$ has intimate connection to the error of CP phase, an appropriate measure in precision measurement era, as discussed in \cite{Machado:2013kya}. It suggests that the CP exclusion fraction can be a universally usable measure for CP sensitivity.
}
In \cite{Machado:2013kya} $f_{\rm CPX}$ was used to display sensitivity to the lepton Kobayashi-Maskawa phase achievable in the ongoing long-baseline neutrino oscillation experiments. For earlier discussion of a CP sensitivity measure closely related to $f_{\rm CPX}$ see \cite{Winter:2003ye,Huber:2004gg}.

In the following two sections, \ref{sec:observables} and \ref{sec:analytic-estimation}, we briefly review the basic features of the observables used and describe simple analytic estimation of the effects of Majorana phases. The expert readers must go directly to the analysis results given in sections~\ref{sec:why} and \ref{sec:results-exclusion-fraction}, starting from describing the analysis method in section~\ref{sec:analysis-method}.

\section{Assumptions and observables used in the analysis}
\label{sec:observables}

In this work we assume that the neutrinos are Majorana particles, 
and their Majorana masses are the unique source of neutrinoless 
double beta decay. 
We also assume the standard three-flavor mixing scheme of neutrinos.

In our analysis we consider the following three observables \cite{Fogli:2004as}: 

\begin{enumerate}

\item
the effective neutrino mass measured by neutrinoless double beta 
(hereafter denoted as $0\nu\beta\beta$) decay experiment
\begin{eqnarray}
m_{0\nu\beta\beta} 
&= &\biggl | m_1 |U_{e1}|^2 + 
m_2 |U_{e2}|^2 \text{e}^{i\alpha_{21}} + 
m_3 |U_{e3}|^2 \text{e}^{i\alpha_{31}} \biggr | \nonumber \\ 
&= &
\biggl | m_1 c_{13}^2 c_{12}^2 + 
m_2 c_{13}^2 s_{12}^2 \text{e}^{i\alpha_{21}} + 
m_3 s_{13}^2 \text{e}^{i\alpha_{31}}  \biggr |, 
\label{eq:0nubb-1}
\end{eqnarray}

\item
the sum of the three neutrino masses 
which can be determined by cosmological observation
\begin{equation}
\Sigma \equiv m_1 + m_2 + m_3,
\label{eq:Sigma-def}
\end{equation}

\item
the effective neutrino mass measured by single beta decay experiment 
\begin{eqnarray}
m_\beta =
\left[ m_1^2\ c_{13}^2 c_{12}^2 + 
m_2^2\ c_{13}^2 s_{12}^2 + m_3^2\ s_{13}^2 \right]^{\frac{1}{2}}, 
\label{eq:mass-beta}
\end{eqnarray}

\end{enumerate}
where $m_i$ ($i=1,2,3$) are neutrino masses and $U_{ek}$ $(k=1,2,3)$ are 
the element of electron-row of the Maki-Nakagawa-Sakata (MNS) 
neutrino mixing matrix \cite{MNS}. 
We use its standard parameterization with notations 
$c_{ij} \equiv \cos \theta_{ij}$, 
$s_{ij} \equiv \sin \theta_{ij}$ \cite{Agashe:2014kda}, 
whereas $\alpha_{21}$ and $\alpha_{31}$ denote 
the Majorana CP phases, in which we have 
redefined $\alpha_{31} - 2 \delta$ in 
Eq. (14.82) in \cite{Agashe:2014kda} as $\alpha_{31}$ for simplicity.

All the three observables above are sensitive to masses of neutrinos, 
but they display different dependences on the mixing parameters, 
which makes them complementary to each other to detect 
the absolute mass scale. 
From the viewpoint of exploring the Majorana phase, $m_{0\nu\beta\beta}$
and the other two play different roles under the assumed errors taken in
our analysis. 
Of course, the information on Majorana phase is 
contained only in $m_{0\nu\beta\beta}$, 
but the constraint on the absolute mass scale from 
the other two is indispensable to produce sensitivity to the phase, 
as we will see clearly in section~\ref{sec:why}. 
The latter two are also complementary to each other. Cosmological measurement of $\Sigma$, though it may achieve greater accuracy in the future, relies on the assumption that we understand the rest of the universe. On the other hand, no such assumption is necessary for beta decay measurement in the laboratories as far as they can reach the neutrino mass scale given by nature. 

In the rest of this section, after briefly mentioning their current
status, we discuss some relevant issues on them in relationship to 
our analysis in this paper.

\subsection{Current status and issues in neutrinoless double beta decay}
\label{subsec:double-beta}

The rate of $0\nu\beta\beta$ decay, or the inverse of the half life time
$T_{1/2}^{0\nu}$ of the $0\nu\beta\beta$ decay, is related to the
effective neutrino mass $m_{0\nu\beta\beta}$ in (\ref{eq:0nubb-1}) as 
\begin{equation}
[T_{1/2}^{0\nu}]^{-1} = G_{0\nu} \left|{\cal{M}}^{(0\nu)}\right|^2 
\left( \frac{m_{0\nu\beta\beta}}{m_e}\right)^2, 
\label{eq:decay-rate}
\end{equation}
where $m_e$ is the electron mass, $G_{0\nu}$ is the kinematic phase space factor, which can be calculated very accurately \cite{Pantis:1996py,Kotila:2012zza}, while ${\cal{M}}^{(0\nu)}$ is the nuclear matrix element (NME) corresponding to the $0\nu\beta\beta$ transition which is largely uncertain. 

The next generation $0 \nu \beta\beta$ decay search aims at 
reaching the region of $m_{0\nu\beta\beta}$ to $\sim 10$ meV, 
which cover the entire inverted mass ordering branch. 
(See Fig.~\ref{fig:m0nbb-m0}.)
Toward approaching to the goal, the ongoing experiments are improving
the lower bound of $0\nu\beta\beta$ decay lifetime. Using $^{76}$Ge
nuclei GERDA obtained the limit  $T_{1/2}^{0\nu} > 2.1 \times 10^{25}$
years at 90\% CL, which is translated into the upper bound on
$m_{0\nu\beta\beta}$ of the range $0.2-0.4$ eV 
depending upon the NME used \cite{Agostini:2013mzu}. 
KamLAND-Zen \cite{Gando:2012zm} and EXO-200 \cite{Albert:2014awa}, both using $^{136}$Xe, derived the bound on lifetime 
$T_{1/2}^{0\nu} > 1.9 \times 10^{25}$ (90\% CL) and 
$T_{1/2}^{0\nu} > 1.1 \times 10^{25}$ (90\% CL) years, respectively. The
latter bound becomes less stringent than the previously reported one
from EXO-200  \cite{Auger:2012ar} because they found 9.9 events
consistent with the $0 \nu \beta \beta$ decay signal. 
A joint analysis of KamLAND-Zen and the previous EXO-200 results are carried out by the KamLAND-Zen group who obtained a bound 
$T_{1/2}^{0\nu} > 3.4 \times 10^{25}$ (90\% CL) \cite{Gando:2012zm}. 
It corresponds to the upper limit of the effective mass $m_{0\nu\beta\beta} < 0.1-0.25\  \ \text{eV\ \ (90\%\ CL)}$. 

The largest uncertainty in translating the bounds on $T_{1/2}^{0\nu}$ to
that of $m_{0\nu\beta\beta}$ comes from the uncertainty of the NME,
${\cal{M}}^{(0\nu)}$. 
Currently, it produces differences of a factor of $\sim 2-4$ on 
$m_{0\nu\beta\beta}$ for a given measured lifetime $T_{1/2}^{0\nu}$. Despite that extensive efforts have been devoted to this issue, there are still considerable differences among NME values obtained by different authors and by different calculation methods. Generally speaking, the difference between different authors is not very large if the same calculation technique is used. The best hope would be to normalize the coupling strength by using the $2\nu$ decay by which the difference may go down to $\simeq 30\%$ level, as proposed within the framework of quasi-particle random phase approximation (QRPA) method \cite{Rodin:2003eb,Rodin:2007fz}. But, notable difference persists to among results calculated by different techniques.\footnote{
It appears that the QRPA
method~\cite{Kortelainen:2007rh,Kortelainen:2007mn,Simkovic:2009pp} and
the interacting boson model~\cite{Barea:2009zza,Barea:2013bz}
systematically 
lead to larger NME values than that calculated by the interacting shell model~\cite{Caurier:2007wq,Menendez:2008jp}. See e.g., Table I and Fig. 1 of Ref.~\cite{Barea:2012zz} in which a comparison between the NME values obtained by these models is made.
}
%
We expect that once the positive signal of $0\nu\beta\beta$ will be observed by using different isotopes, by comparing the real data from these different isotopes, it would be possible to check the validity of various theoretical calculations based on different models or methods in a systematic way, which should allow to reduce significantly the NME uncertainty. For this reason we consider uncertainty on the NME a factor of less than or equal to 2 in this paper. 

Recently, it was pointed out that renormalization effect on the axial
vector coupling constant in the nuclear medium can have a sizable impact
on the relationship between $m_{0\nu\beta\beta}$ and half life
\cite{Barea:2013bz}. 
It is likely that it affects the analysis which aims at detecting 
the effects of the Majorana phase such as ours in an $A$-dependent
manner,  and it is important to discuss influence of the effect.  
Unfortunately, we cannot address this issue in this paper because of 
the simplicity of our $\chi^2$ construction. 

\subsection{Cosmological observation of sum of neutrino masses}
\label{subsec:sum-mass}

The recent high precision cosmological observation became sensitive to 
the sum of the neutrino masses $\Sigma$ defined in
(\ref{eq:Sigma-def}). 
With baryon acoustic oscillation (BAO) data the Planck collaboration 
obtained their most stringent limit 
$\Sigma < 0.23 \ \text{eV}~(\text{Planck + WMAP + highL + BAO})$ 
at 95\% CL~\cite{Ade:2013zuv}. 
The authors of Ref.~\cite{Riemer-Sorensen:2013jsa} obtained 
a stronger bound, $\Sigma < 0.18 \ \text{eV}$ at 95\% CL, 
by adding the large-scale matter power spectrum data from 
the WiggleZ Dark Energy Survey \cite{Parkinson:2012vd}.

More recently, some analyses started to favor non-zero best fit value of
$\Sigma$ ~\cite{Battye:2013xqa,Wyman:2013lza,Hamann:2013iba} of active
or sterile neutrinos. There exists some tension between strength of
cluster correlations determined by CMB and the lensing observations
within the Planck data itself as well as including other data sets. One
of the ways to reduce the tension is to introduce neutrino masses. The
authors of \cite{Battye:2013xqa}, 
assuming the active neutrinos, 
obtained the best fit value $\Sigma = 0.32 \pm 0.081$ eV, 
which implies that $m_1 \sim m_2 \sim m_3 \sim 0.1$ eV in 
the three flavour scheme. 
We expect that the issue is to be resolved in future 
cosmological observation; 
The data either from galaxy surveys or weak lensing by 
the Euclid satellite, with Planck constraints, 
may lead to the sensitivity to 
$\Sigma \simeq 0.01-0.05$ eV~\cite{Carbone:2010ik,Kitching:2008dp,Amendola:2012ys,Hamann:2012fe}. 

Despite the progress in tightening up the neutrino mass bound by cosmological observation and the great precision expected in the future, we should keep in mind that the analysis has done and needs to be done within a particular cosmological model. Since neutrinos plays relatively minor role in determining the CMB spectrum and affecting the structure formation, we suspect that precision measurement of $\Sigma$ requires a well established framework describing our universe and its time evolution.\footnote{
The minor role played by neutrinos in the cosmos is indicated, for example, by the fact that the best fit to the current cosmological data does not require presence of neutrino masses, as reported in ~\cite{Leistedt:2014sia}. 
}
Therefore, as a prerequisite of our analysis in this paper, we assume that the {\em standard model (SM) of cosmology} is established at the time one can execute precision measurement of half-life of $0\nu\beta\beta$ decay. It would not be the too optimistic to assume it because now already precise cosmological data by the Planck and the other observations are well described by the $\Lambda$CDM model (see Sec.~\ref{subsec:sum-mass}). 

\subsection{Measurement of energy spectrum near the end point in single beta decay}
\label{subsec:mass-beta}

The absolute neutrino mass scale can be probed by laboratory experiment 
which measures in a high precision electron energy spectrum at the end
point in single beta decay \cite{Otten:2008zz}. 
The effective neutrino mass $m_\beta$ measured by this method can be 
written as in (\ref{eq:mass-beta}). 
Currently, the best upper bound on $m_\beta$ comes from 
Mainz~\cite{Kraus:2004zw} and Troitsk~\cite{Aseev:2011dq} experiments 
which place the upper bound $m_\beta < 2.3$ eV  and $m_\beta < 2.05$ eV, 
respectively, both at 95\% CL.

In the near future, we expect that the bound will be improved by an order of magnitude by the KATRIN experiment \cite{Wolf:2008hf} which is expected to reach the sensitivity of $m_\beta \sim$ 0.2 eV at 90\% CL. If the neutrino mass pattern is indeed the almost degenerate type, as suggested by some analyses of recent cosmological data \cite{Battye:2013xqa,Wyman:2013lza,Hamann:2013iba}, KATRIN has a chance of executing first model-independent determination of the absolute neutrino mass, and at the same time also serves for cross checking the cosmological measurement.

\section{Analytic estimate of the effect of Majorana phase} 
\label{sec:analytic-estimation} 

In this section, we discuss qualitative features of effects of the Majorana phase onto the $0\nu\beta\beta$ decay observable $m_{0\nu\beta\beta}$, aiming at illuminating how and where the best sensitivity to the Majorana phase can be expected. See also the discussions in the references cited in Sec.~\ref{sec:intro}, as well as, e.g., in Refs.~\cite{Pascoli:2002xq,Pascoli:2002ae,Petcov:2004wz,Lindner:2005kr}. After presenting a familiar (and informative) plot of $m_{0\nu\beta\beta}$ as a function of $m_0$ \cite{Vissani:1999tu}, the lowest neutrino mass, we start by mentioning a general properties of $m_{0\nu\beta\beta}$.

\subsection{Allowed regions of $m_{0\nu\beta\beta}$ as a function of lightest neutrino mass}
\label{subsec:current-range}

In this paper we use the lightest neutrino mass (rather than $\Sigma$), 
denoted as $m_0$, as the relevant parameter (observable) to be 
determined by the experiments. 
Namely, $m_0 \equiv m_1$ and $m_0 \equiv m_3$ for the normal and 
the inverted mass orderings, respectively. 
All the other masses and their differences can be 
calculated by using the relations 
\begin{eqnarray}
m_1 \equiv m_0, \ 
m_2 = \sqrt{m_0^2 + \Delta m^2_{21} }\ , \ 
m_3 = \sqrt{m_0^2 + \Delta m^2_{21} + \Delta m^2_{\text{32}} }
\ \text{(normal mass ordering)}, 
\label{eq:mass-normal}
\\
m_1 = \sqrt{m_0^2 - \Delta m^2_{21} -\Delta m^2_{32} }\ , \ 
m_2 = \sqrt{m_0^2 -  \Delta m^2_{\text{32}} }\ , \ 
m_3 \equiv m_0 
\ \text{(inverted mass ordering)},
\label{eq:mass-inverted}
\end{eqnarray}
%
%
\begin{figure}
\hglue  0.4cm
\includegraphics[width=1.00\textwidth]{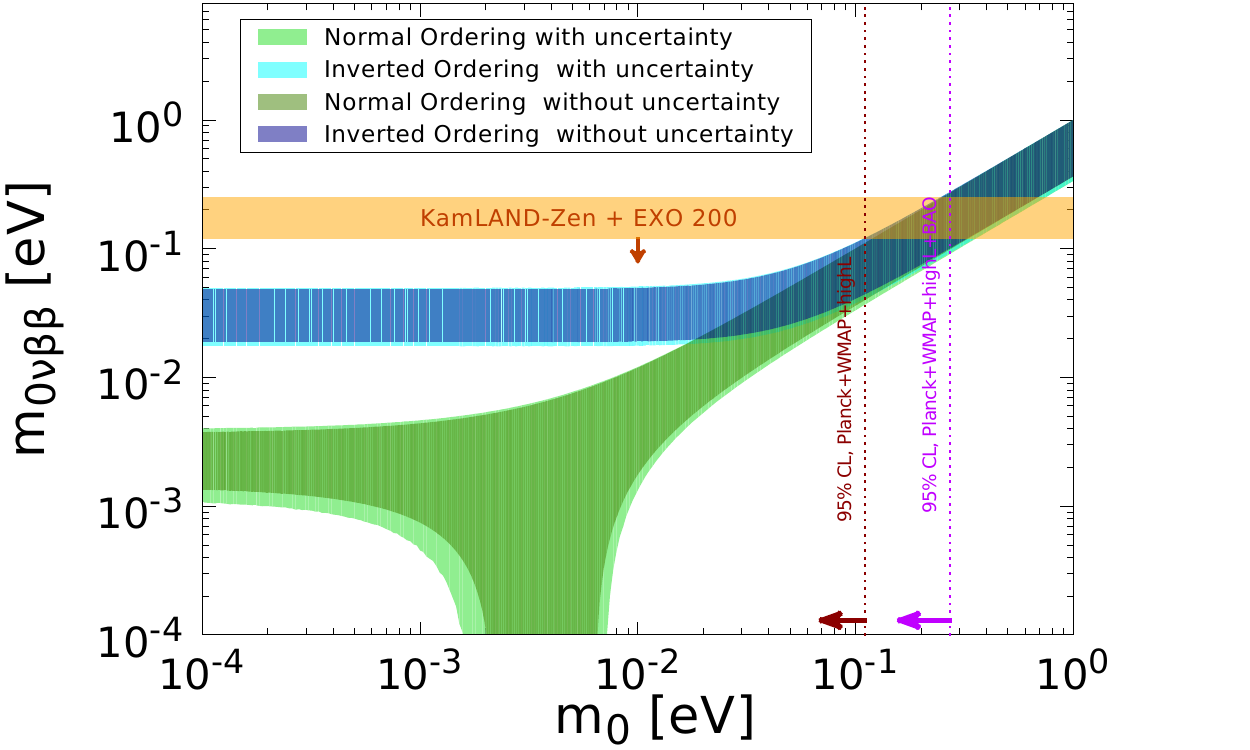}
\vglue -0.3cm
\caption{
The currently allowed ranges of $m_{0\nu\beta\beta}$ 
observables of $0\nu\beta\beta$ decay is shown 
as a function of the lightest neutrino mass $m_0$. 
In the case of normal (inverted) mass ordering the ranges are 
shown by green (blue) colour. 
The light (dark) coloured regions are computed by taking into account 
(without taking account) the current 1$\sigma$
uncertainties of the relevant 
mixing parameters.
Also shown are the limits on $m_{0\nu\beta\beta}$ 
coming from KamLAND-Zen and 
EXO~\cite{Gando:2012zm} (by the light brown band and arrow), 
the bounds on $m_0$ obtained by 
Planck collaboration~\cite{Ade:2013zuv} 
(by the magenta and the brown dotted lines).
We note that the KamLAND-EXO bound spans a band (not line) because of the NME uncertainty. 
}
\label{fig:m0nbb-m0}
\end{figure}
%
In Fig.~\ref{fig:m0nbb-m0}, we show the allowed region of 
$m_{0\nu\beta\beta}$ as a function of $m_0$ for both the normal and 
the inverted mass orderings. The allowed region is calculated by the two
ways, without and with the current 1$\sigma$ uncertainties of the mixing 
parameters, which are indicated, respectively, by the dark and light
colors, with the blue and green colors for the inverted and normal mass 
ordering. 
It is remarkable that the effect of 1$\sigma$ uncertainties of the mixing parameters is quite small by now. In contrast, variation over the Majorana phases $\alpha_{21}$ and $\alpha_{31}$ gives much larger impact on allowed region of $m_{0\nu\beta\beta}$, not only producing sizeable width but also creating a down-going branch in region $10^{-3}\ \text{eV} \lsim m_0 \lsim 10^{-2}\ \text{eV}$ for the case of the normal mass ordering due to the strong cancellation of the three mass terms in Eq.~(\ref{eq:0nubb-1}).

The difference between the normal and the inverted mass ordering becomes 
clearly visible only in small $m_0$ region, typically $m_0 \lsim 0.1
\text{eV}$.
Therefore, onset to the almost degenerate mass regime, which is formally 
defined as $m_0 \gg \sqrt{ \Delta m^2_{\text{atm}} } \simeq 0.05$ eV, 
already takes place for $m_{0\nu\beta\beta}$ and $m_{\beta}$ at around $m_0 \gsim 0.1 \text{eV}$.

In Fig.~\ref{fig:m0nbb-m0} and in the rest of this paper we use the
values obtained in \cite{Capozzi:2013csa}: 
\begin{eqnarray}
\Delta m^2_{21} &=  & 7.54 \times 10^{-5}\ \text{eV}^{2}, \nonumber \\
\sin^2\theta_{12} & = & 0.308\ \text{or}\ \sin^2 2\theta_{12} = 0.853,
\label{eq:best-fit1}
\end{eqnarray}
for the both mass orderings whereas 
\begin{eqnarray}
\Delta m^2_{32} &=  & 2.40\ (-2.44)\times 10^{-3}\ \text{eV}^{2}, \nonumber \\
\sin^2\theta_{13} & = & 0.0234\ (0.0239) \ 
\text{or} \ \sin^2 2\theta_{13} = 0.0914\ (0.0933), 
\label{eq:best-fit2}
\end{eqnarray}
for the normal (inverted) mass ordering.\footnote{
Since the best fitted values of mass squared difference for 
atmospheric neutrino oscillation, $\Delta m^2$, shown in 
Table I of Ref.~\cite{Capozzi:2013csa}, 
is defined as $\Delta m^2 \equiv m^2_3-(m_1^2+m_2^2)/2$, 
the values of $\Delta m^2_{32}$ in Eq.~(\ref{eq:best-fit2})
were obtained by using the relation
$\Delta m^2_{32} = \Delta m^2 - \Delta m^2_{21}/2$.
}
%
Using these values of the parameters $m_{0\nu\beta\beta}$ at its current status can be summarized as
\begin{eqnarray}
m_{0\nu\beta\beta} 
\simeq
\left| 0.676\,(0.675)\ m_1 + 
0.301\ m_2 \text{e}^{i\alpha_{21}} + 
0.0234\,(0.0239)\ m_3 \text{e}^{i\alpha_{31}}  \right|
\label{eq:0nubb-3}
\end{eqnarray}
for the normal (inverted) mass ordering.

\subsection{Reduction of Majorana phase space}
\label{subsec:reduction}

We describe here a general properties of $m_{0\nu\beta\beta}$ given in (\ref{eq:0nubb-1}). Since the phases $\alpha_{21}$ and $\alpha_{31}$ enter into $m_{0\nu\beta\beta}$ in the form of either one of $\cos \alpha_{21}$, $\cos \alpha_{31}$, and $\cos (\alpha_{31} - \alpha_{21})$ $m_{0\nu\beta\beta}$ has the following reflection symmetry 
\begin{equation}
m_{0\nu\beta\beta}(m_0,\alpha_{21}, \alpha_{31}) = 
m_{0\nu\beta\beta}(m_0,2\pi-\alpha_{21},2\pi- \alpha_{31}).
\label{eq:symmetry}
\end{equation}
Notice that it is an invariance under the simultaneous transformation of two phase variables $\alpha_{21}$ and $\alpha_{31}$, and each one of them alone does not keep $m_{0\nu\beta\beta}$ invariant.

\subsection{The case of almost degenerate mass  spectra}
\label{subsec:degenerate}

The best place to discuss clearly how and where the best sensitivity to the Majorana phase can be expected is in the almost degenerate regime, $m_1 \approx m_2 \approx m_3$, as ensured if $m_0 \gg \sqrt{\Delta m^2_{\text{atm}}} \simeq 0.05$ eV, where $\Delta m^2_{\text{atm}} \equiv  |\Delta m^2_{32} |$. Moreover, the regime turns out to be the main target of our analysis, given reasonable values of experimental errors in the next generation $0\nu\beta\beta$ decay experiments.
In this regime, the effective mass $m_{0\nu\beta\beta}$ in (\ref{eq:0nubb-1}) can be written approximately to leading order in $s^2_{13}$ as 
\begin{eqnarray}
m_{0\nu\beta\beta} \simeq c_{13}^2 m_0 
\times \left[ 
1 - \sin^2{2 \theta_{12}} 
\sin^2\left(\frac{\alpha_{21}}{2} \right) 
+ \frac{1}{2}\sin^2 2\theta_{13}
\left\{ c_{12}^2 \cos \alpha_{31} 
+ s_{12}^2 \cos(\alpha_{21}-\alpha_{31})\right\}
\right]^{\frac{1}{2}} 
\label{eq:0nubb-degenerate}
\end{eqnarray}
for the both mass orderings. The dominant effect of the Majorana phase is induced by $\alpha_{21}$ because the last term in the square bracket in (\ref{eq:0nubb-degenerate}) has a $\theta_{13}$ suppression and is smaller by a factor of $10-20$ than the other terms. Ignoring the $\sin^2 2\theta_{13}$ term, the maximum  and minimum values of $m_{0\nu\beta\beta}$ (upper and lower edges of  the bands for $m_0 \gsim 0.1$ eV in the plot in  Fig.~\ref{fig:m0nbb-m0}) are simply given, respectively, by 
\begin{eqnarray}
m_{0\nu\beta\beta}^{\text{max}} 
\simeq c_{13}^2 m_0, \ \ 
m_{0\nu\beta\beta}^{\text{min}} 
\simeq 
c_{13}^2 m_0 \cos 2 \theta_{12} 
\simeq 0.383~m_{0\nu\beta\beta}^{\text{max}}.
\end{eqnarray}
It means that the ratio of the maximum to the minimum value of effective mass changes by a factor of $m_{0\nu\beta\beta}^{\text{max}} / m_{0\nu\beta\beta}^{\text{min}} \simeq 2.6$ as the Majorana phase is varied. Therefore, if one can determine $m_{0\nu\beta\beta}$ from experiment with errors smaller than the factor 2.6, after taking into account the uncertainty coming from the NME, it is in principle possible to constrain the Majorana phase $\alpha_{21}$.

Since the sensitivity to the Majorana phase arises in places where the NME uncertainty cannot obscure the effect of phases, the best place is at the maximal and the minimal value of $m_{0\nu\beta\beta}$. One can show that the best sensitivity to $\alpha_{21}$ is obtained at either $\alpha_{31} = 0$ or $\pi$.\footnote{
$m_{0\nu\beta\beta}$ in (\ref{eq:0nubb-degenerate}) is nearly periodic 
function of $\alpha_{21}$ and $\alpha_{31}$, and has 
the following properties in regions $0 \le \alpha \le \pi$: 
(1) The function inside parenthesis in Eq.~(\ref{eq:0nubb-degenerate})
is an extremely slowly decreasing function of $\alpha_{31}$ for in the
entire region of $\alpha_{21}$. (2) It is a decreasing function of
$\alpha_{21}$ in the entire region of $\alpha_{31}$, and varies from
$\simeq 1$ to $\simeq 0.15$ as $\alpha_{21}$ is varied 
from $0$ to $\pi$. Therefore, it has the maximum (minimum) at $\alpha_{21} =
\alpha_{31} = 0$ ($\alpha_{21} = \alpha_{31} =\pi$). 
}
On the contrary, the worst sensitivity is expected at around the mean
value of $m_{0\nu\beta\beta}$ between their maximum and minimum
values. Typically, they are at around $\alpha_{21} \simeq {2\pi}/{3}$
for the whole range of $\alpha_{31}$. These features will be confirmed
explicitly by quantitative analysis in section 6.

\subsection{Hierarchical case; inverted mass spectrum}

Now let us consider the case of $m_0 < \sqrt{\Delta m^2_{21}} \lsim
0.01$ eV for the inverted mass ordering. 
In this case, $m_1 \approx m_2 \sim \sqrt{\Delta m^2_{\text{atm}}} \gg m_3 \equiv m_0$. This is the regime corresponding to the horizontal band for the inverted mass spectra in Fig.~\ref{fig:m0nbb-m0}. It is interesting to observe that essentially the same treatment can go through for $m_{0\nu\beta\beta}$ as in the case of almost degenerate spectrum. In the case of strongly hierarchical inverted mass spectrum the $m_3$ term in (\ref{eq:0nubb-3}) is doubly suppressed than other two terms by a factor of ${m_3}/{m_2} \ll [\Delta m^2_{21} / \Delta m^2_{\text{atm}} ]^{1/2} \simeq 0.18$ and $s^2_{13} = 0.0234\,(0.0239)$ for the normal (inverted) ordering. In the case of almost degenerate spectrum the suppression of $m_3$ term is only due to the latter factor.
Under the approximation of ignoring the $m_3$ term, $m_{0\nu\beta\beta}$ is expressed as in the degenerate case, (\ref{eq:0nubb-degenerate}) without $s^2_{13}$ term, but replacing $m_0$ by $\sqrt{\Delta m^2_{\text{atm}}}$, 
\begin{eqnarray}
m_{0\nu\beta\beta} \simeq
c_{13}^2  \sqrt{ \Delta m^2_{\text{atm}} } 
\left[ 1 - \sin^2{2 \theta_{12}} 
\sin^2\left(\frac{\alpha_{21}}{2} \right) \right]^{\frac{1}{2}}.
\label{IH-hierarchical}
\end{eqnarray}
Since the dependence of $m_{0\nu\beta\beta}$ on the Majorana phase is essentially the same as in the case of the degenerate mass regime, the same results hold for the sensitivity to $\alpha_{21}$ and the location of highest sensitivity. Under our assumption of the precision of measurement, $\sigma_{0\nu\beta\beta} (\equiv$ error of $m_{0\nu\beta\beta}) = 0.01$ eV (see section~\ref{sec:analysis-method}), however, the sensitivity to $\alpha_{21}$ is quite limited in this regime when it is combined with the NME uncertainty. It will be demonstrated in the analysis in section~\ref{sec:results-exclusion-fraction}.

\subsection{Hierarchical case; normal mass spectrum}

If $m_0 \ll \sqrt{\Delta m^2_{21}}$, typically $m_0 \sim 10^{-4}$ eV,
one can ignore $m_1 (= m_0)$ term in Eq.~(\ref{eq:0nubb-3}). 
Then, we obtain the expression of $m_{0\nu\beta\beta}$ as 
\begin{eqnarray}
m_{0\nu\beta\beta} & = &
\sqrt{ \Delta m^2_{21} } 
\left[ 
s^4_{12} c^4_{13} + 
 2 \sqrt{ \frac{ s^4_{13} }{ \epsilon } } 
 s^2_{12} c^2_{13} \cos ( \alpha_{31} - \alpha_{21} )
\right]^{1/2} \nonumber \\
&\simeq &
\sqrt{ \Delta m^2_{21} } 
\left[ 
0.0905 +  0.0794\ \cos ( \alpha_{31} - \alpha_{21} )
\right]^{1/2} 
\label{mee-normal}
\end{eqnarray}
where $\epsilon \equiv
\Delta m^2_{21}/\Delta m^2_{\text{atm}} \simeq 0.0314$ 
and we have kept order 
$\sqrt{{s^4_{13}}/{\epsilon}}$ 
term which is 
$\simeq 0.13$, but ignored the terms of order $\epsilon^2$, $s^4_{13}$, 
${s^4_{13}}/{\epsilon} \simeq {5.3 \times 10^{-4}}/{0.0314} \simeq 0.017$ 
and $\sqrt{\epsilon} s^2_{13} \simeq 4.1 \times 10^{-3}$. 
Thus, in the normal mass ordering with small $m_0$ region, 
the phase $\alpha_{31}$ (in a combination $\alpha_{31} - \alpha_{21}$) affects the $0\nu\beta\beta$ decay rate, 
in contrast to the case of almost degenerate and inverted
mass spectra in which only $\alpha_{21}$ plays a role.
As $\alpha_{31}$ changes $m_{0\nu\beta\beta}$ varies as 
\begin{eqnarray}
\sqrt{ \Delta m^2_{21} } 
\left[ s^4_{12} c^4_{13} - 
 2 \sqrt{ \frac{ s^4_{13} }{ \epsilon } }  s^2_{12} c^2_{13} \right]^{1/2} 
\leq m_{0\nu\beta\beta} 
\leq \sqrt{ \Delta m^2_{21} } 
\left[ s^4_{12} c^4_{13} + 
 2 \sqrt{ \frac{ s^4_{13} }{ \epsilon } }  s^2_{12} c^2_{13} \right]^{1/2}. 
\label{mee-range-NH}
\end{eqnarray}
Or numerically, 
\begin{eqnarray}
0.105 \sqrt{ \Delta m^2_{21} } \simeq 9.12 \times 10^{-4}\ \text{eV}
\leq m_{0\nu\beta\beta} 
\leq 
0.412 \sqrt{ \Delta m^2_{21} } \simeq 3.58 \times 10^{-3}\ \text{eV} 
\label{mee-range-NH2}
\end{eqnarray}
a factor of $\sim 4$ variation. 
Again, under our assumption of the precision of 
$\sigma_{0\nu\beta\beta} = 0.01$ eV
it would be very difficult to probe the value of $\alpha_{31}$ in this regime, as we will see in section~\ref{sec:results-exclusion-fraction}.

\section{Analysis Method}
\label{sec:analysis-method}

\subsection{Assumptions on experimental errors}
\label{sec:exp-errors}

We use the three observables, sum of neutrino masses $\Sigma$, 
which is determined by cosmology, 
and the effective mass parameters 
$m_\beta$ and $m_{0\nu\beta\beta}$ which are measured,
respectively, by single and double beta decay experiments. 
Let us assume that they are measured with some uncertainties 
around the central values, $m_{0\nu\beta\beta}^{(0)}$, $\Sigma^{(0)}$, and $m_\beta^{(0)}$:
\begin{equation}
m_{0\nu\beta\beta}^{\text{obs}} = 
m_{0\nu\beta\beta}^{(0)}
\pm \sigma_{0\nu\beta\beta},
\label{eq:uncertainty-bb}
\end{equation}
\begin{equation}
\Sigma^{\text{obs}} = \Sigma^{(0)} \pm \sigma_\Sigma,
\label{eq:uncertainty-S}
\end{equation}
\begin{equation}
m_\beta^{\text{obs}} = m_\beta^{(0)} \pm \sigma_\beta,
\label{eq:uncertainty-b}
\end{equation}
where $\sigma_{0\nu\beta\beta}$, $\sigma_\Sigma$, and $\sigma_\beta$ 
denote  the corresponding 1$\sigma$ uncertainties. 
Due to the uncertainty in the NME calculations, one must pay 
special attention to the precise meaning of $m_{0\nu\beta\beta}^{(0)}$.
See the next subsection regarding this point.
We take the following values for the uncertainties,
\begin{equation}
\sigma_{0\nu\beta\beta} = 0.01 \ \text{eV}, \ \
\sigma_\Sigma = 0.05 \ \text{eV}, \ \ 
\sigma_\beta = 0.06 \ \text{eV}, 
\label{reference-error}
\end{equation}
which will be used as the reference values in our analysis. 

Here is some reasonings for our choices of the uncertainties in (\ref{reference-error}).
To have a reasonable estimate of the uncertainty in
$m_{0\nu\beta\beta}$, 
we need a detailed discussion because we must address 
the question of how the experimental uncertainty on $T_{1/2}^{0\nu}$ is translated into that of $m_{0\nu\beta\beta}$, and the treatment heavily depend on to what extent the experimental backgrounds can be suppressed. We present such a discussion in Appendix ~\ref{sec:double-beta-sensitivity}, based on the earlier treatment e.g., in Refs.~\cite{Elliott:2002xe,Cremonesi:2013bma}.
Under the assumption of dominance of statistical error, we will argue there that the uncertainty of 0.01 eV may be reachable in future experiments that are upgraded to a ton scale. Such experiments will be able to cover the entire allowed range corresponding to the inverted mass ordering, and the similar region of $m_{0\nu\beta\beta}$ in the normal mass ordering. 

As is mentioned in Sec.~\ref{subsec:sum-mass}, we assume that the SM of cosmology will be established by the time the $0\nu\beta\beta$ decay experiments reach the above accuracy. Because of the success of the $\Lambda$CDM model to explain the current data sets it is likely that the SM of cosmology is not too far away from the $\Lambda$CDM model. Then, we may be allowed to use the value of the uncertainty in $\Sigma$ estimated within the current framework of $\Lambda$CDM model. It is expected that the future galaxy surveys \cite{Carbone:2010ik} and weak lensing \cite{Kitching:2008dp} both lead, under the Planck constraint, the sensitivity to $\Sigma$ to the level of $0.02 - 0.05$ eV. According to the authors of Ref.~\cite{Hamann:2012fe}, in the most optimistic case, the sensitivity could even lead to $\sim 0.01$ eV. Therefore, our reference error on $\Sigma$ would be in a conservative side, and the alternative choice $\sigma_\Sigma= 0.02$ eV which we will also examine would not be too optimistic in view of \cite{Hamann:2012fe}. 

For KATRIN \cite{KATRIN} they quote the error 0.025 eV$^2$ for $m_{\beta}^2$, which may be translated into the error of 0.063 eV for $m_{\beta}$ for the case where true value of $m_0=0.2$ eV, which justifies our choice in (\ref{reference-error}).

\subsection{Analysis procedure: $\chi^2$ and treatment of NME errors}
\label{sec:analysis-procedure}

As we emphasized in section~\ref{subsec:double-beta} the error of $m_{0\nu\beta\beta}$ due to uncertainties of the NME would be a major obstacle to extract informations of the Majorana phases from $0\nu\beta\beta$ decay experiments. We take into account the uncertainties of NME via the following way, which is very similar to the one proposed in \cite{Pascoli:2005zb} but with different functional from. 
Let us assume that the unknown true value of NME, which we denote as ${\cal{M}}^{(0\nu)}$, exists in a range ${\cal{M}}_{{\text{min}}}^{(0\nu)} \le {\cal{M}}^{(0\nu)} \le {\cal{M}}_{{\text{max}}}^{(0\nu)}$. 
Practically, ${\cal{M}}_{{\text{min}}}^{(0\nu)}$ and
${\cal{M}}_{{\text{max}}}^{(0\nu)}$ imply, 
respectively, the minimum and the maximum values of NME 
calculated in a consistent framework. We may define the NME uncertainty factor as 
$r_{\text{\tiny NME}} \equiv 
{\cal{M}}_{{\text{max}}}^{(0\nu)}/
{\cal{M}}_{{\text{min}}}^{(0\nu)}$. 
Then, if we define $\xi$ as 
\begin{equation}
\xi \equiv 
\frac{ {\cal{M}}_{{\text av}}^{(0\nu)} }{ {\cal{M}}^{(0\nu)} } 
\label{eq:xi}
\end{equation}
where ${\cal{M}}_{{\text av}}^{(0\nu)} \equiv 
\left( {\cal{M}}_{{\text{max}}}^{(0\nu)}
       {\cal{M}}_{{\text{min}}}^{(0\nu)} 
\right)^{1/2}$, 
$\xi$ falls into a region 
$1/\sqrt{r_{\text{\tiny NME}}} \le \xi \le \sqrt{r_{\text{\tiny
NME}}}$. We note that our $\xi$ corresponds to inverse of $\xi$ 
defined in \cite{Pascoli:2005zb}.

In this work, we consider the three values of the NME uncertainty
parameter, $r_{\text{\tiny NME}}$ = 1.3, 1.5 and 2.0. 
(Exceptional use of an extreme value, $r_{\text{\tiny NME}}$ = 1.1, is
made at the end of section~\ref{sec:results-exclusion-fraction}.) 
We note that, strictly speaking, errors associated with the three 
observables we consider in this work would not be  Gaussian, in
particular, the one from cosmology. 
It should be better to do analysis based on Monte Carlo simulations, 
as performed, e.g., in Refs.~\cite{Host:2007wh,Hannestad:2007tu}. 
However, we believe that the procedure we use produces approximately 
correct results, which are sufficient for our purpose. 

Using the three observables in 
(\ref{eq:uncertainty-bb}) - (\ref{eq:uncertainty-b}), 
we define the $\chi^2$ function as follows, 
\begin{eqnarray}
\chi^2   \equiv
{\text{min}} 
\left\{
\left[
\frac{\Sigma^{(0)}-\Sigma^{\text{fit}}(m_0)}{\sigma_\Sigma}
\right]^2 
\hskip -0.1cm + \hskip -0.1cm
\left[\frac{m_\beta^{(0)}-m_\beta^{\text{fit}}(m_0)
} {\sigma_\beta}\right]^2 \right. 
\hskip -0.2cm + \hskip -0.1cm
\left. \left[\frac{\xi \  m_{0\nu\beta\beta}^{(0)}-
m_{0\nu\beta\beta}^{\text{fit}}(m_0,\alpha_{21},\alpha_{31})
} 
{\sigma_{0\nu\beta\beta}}
\right]^2  \right\}, 
\label{eq:chi2}
\end{eqnarray}
%
where $\Sigma^{(0)}$, $m_\beta^{(0)}$ and $m_{0\nu\beta\beta}^{(0)}$ are the central values to be determined by the future experiments whereas $\Sigma^{\text{fit}}(m_0)$, $m_\beta^{\text{fit}}(m_0)$ and $m_{0\nu\beta\beta}^{\text{fit}}(m_0,\alpha_{21},\alpha_{31})$ are the ones to be fitted by varying the parameters $m_0$, $\alpha_{21}$ and $\alpha_{31}$. We also vary the parameter $\xi$ in the interval of $[1/\sqrt{r_{\text{\tiny NME}}}, \sqrt{r_{\text{\tiny NME}}}]$ by taking flat prior assuming no preference among the estimated values of NME. In this way, the NME uncertainty is treated in a similar way as the flux ratio parameter in the solar neutrino analysis to determine the true neutrino flux in comparison to the calculated flux by the Standard Solar Model, as was done, e.g., in Ref.~\cite{Bahcall:2003ce}.


To summarize, we minimize the $\chi^2$ by varying $m_0$, $\alpha_{21}$, $\alpha_{31}$ and $\xi$, and determine the allowed parameter space by imposing the condition, 
\begin{equation}
\Delta \chi^2 \equiv \chi^2 - \chi^2_{{\text{min}}} 
< 2.3, 6.18\ \text{and} \ 11.83 \ 
(1, 4\ \text{and} \ 9)
\end{equation}
for 1, 2 and 3$\sigma$ CL for two (one) degrees of freedom. 

Throughout this paper we use the mixing parameters described in (\ref{eq:best-fit1}) and (\ref{eq:best-fit2}). We do not consider the uncertainties of the mixing parameters apart from the currently unknown mass ordering, because the impact is rather small which is expected to become even smaller in the near future. This is because we expect significant improvement in the accuracies of these parameters by future oscillation experiments, in particular, by the medium baseline ($\sim 50-60$ km) reactor experiments such as JUNO~\cite{Kettell:2013eos} and RENO-50~\cite{Seo:2013swa} as demonstrated, e.g., in Refs.~\cite{Minakata:2004jt,Bandyopadhyay:2004cp,Ge:2012wj,Capozzi:2013psa}.

\section{Sensitivity to the Majorana Phase: Why so much constraint?}
\label{sec:why}

In this section, we have a glance over the global features of sensitivity to the Majorana phase, and try to understand the reasons why such nontrivial constraint on the phase arises despite that a large uncertainty exists in calculating the NME.

\subsection{Allowed regions of $m_0$, $\alpha_{21}$ and $\alpha_{31}$}
\label{sec:results-allowed-regions}

\begin{figure}[!h]
\vglue -8mm
\begin{center}
\hglue 1.0cm
\includegraphics[width=0.95\textwidth]{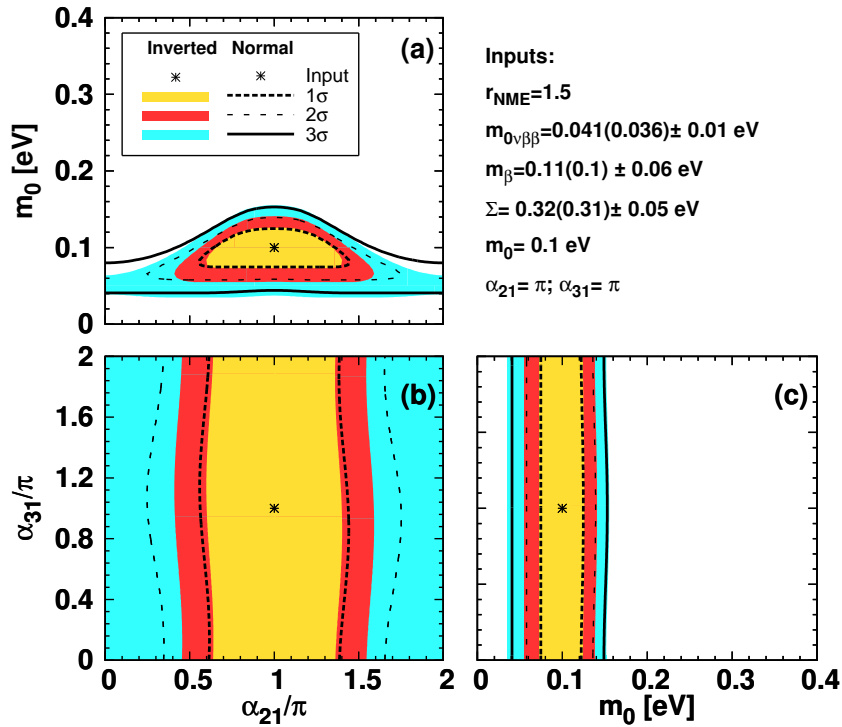}
\end{center}                          
\vglue -1.0cm
\caption{
Regions allowed at 1, 2 and 3$\sigma$ CL indicated by the filled color of yellow, red and light blue, respectively, for 2 DOF, 
projected into the planes of 
(a) $\alpha_{21}-m_0$ (b) $\alpha_{21}-\alpha_{31}$ and 
(c) $m_0-\alpha_{31} $ 
for the case of the inverted mass ordering 
and the true values of $m_0$ = 0.1 eV,  $\alpha_{21} =\alpha_{31}=\pi$, and $r_{\text{\tiny NME}} = 1.5$. 
The allowed contours of the normal mass ordering are shown by the
 dotted (1$\sigma$), 
dashed (2$\sigma$), and solid (3$\sigma$) curves. 
In the legend, the numbers outside (inside) the parentheses 
corresponds to the inverted (normal) mass ordering. 
}
\label{fig:allowed-regions-0.1eV}
\end{figure}

Let us consider the case where the true values of relevant fundamental parameters are, $m_0 = 0.1$ ($\Sigma \simeq 0.32$ eV) and $\alpha_{21} = \alpha_{31} = \pi$. In Fig.~\ref{fig:allowed-regions-0.1eV} the allowed regions projected into the planes of (a) $\alpha_{21}-m_0$ (b) $\alpha_{21}-\alpha_{31}$ and (c) $m_0-\alpha_{31}$ are drawn. The NME uncertainty is taken as $r_{\text{\tiny NME}} = 1.5$. 
In the case of inverted mass ordering they are depicted by the filled colours, yellow, red, and light blue, corresponding, respectively, to 1, 2 and 3$\sigma$ CL for 2 degree of freedom (DOF). While for the normal mass ordering they are drawn by the black dotted (1$\sigma$), dashed ($2\sigma$) and the solid (3$\sigma$) curves. 

We notice from Fig.~\ref{fig:allowed-regions-0.1eV} that it is possible
to restrict $\alpha_{21}$ in some interval at 2$\sigma$ CL for 2 DOF,
but not at 3$\sigma$ CL, the feature which is valid for the both mass
orderings. On the other hand, it is not possible to constrain at all the
phase $\alpha_{31}$, as it can be expected by 
the smallness of the third term proportional to $m_3$ in
(\ref{eq:0nubb-1}) due to $\theta_{13}$ suppression. 
We have checked that this is true for 
any values of $\alpha_{31}$ because 
the dependence of the allowed regions on $\alpha_{31}$ is always very mild. 
We also note that the allowed regions in Fig.~\ref{fig:allowed-regions-0.1eV}(b) are symmetric with respect to $\alpha_{21} = \pi$ in a good approximation. It stems from the simultaneous reflection symmetry $\alpha_{21}  \rightarrow 2\pi - \alpha_{21}$ and $\alpha_{31}  \rightarrow 2\pi - \alpha_{31}$  in Eq.~(\ref{eq:symmetry}) and the weak dependence of $\Delta \chi^2$ on $\alpha_{31}$. 

If we repeat the same exercise with $m_0 = 0.2$ eV ($\Sigma \simeq 0.61$
eV), the $3 \sigma$ allowed region of $\alpha_{21}$ is restricted to
$0.5 \pi \lsim \alpha_{21} \lsim 1.5 \pi$. The difference between the
normal and inverted mass orderings, which is visible at $m_0 \simeq 0.1$
eV (Fig.~\ref{fig:allowed-regions-0.1eV}), diminishes at $m_0 \gsim 0.2$
eV. The difference is evident at $m_0 \lsim 0.05$ eV (not shown here, 
but see \cite{Minakata:2014jba}), where the sensitivity to $\alpha_{21}$ still persists in the case of inverted ordering, as seen in (\ref{IH-hierarchical}). Since the dependence on $\alpha_{31}$ is quite mild, from now on we will show only the allowed regions projected onto the plane of $\alpha_{21}-m_0$, which is the most interesting combination for our purpose. 

\subsection{Why and how does constraint on the Majorana phases arise?} 
\label{how-constrained}

One may ask why such nontrivial constraints on $\alpha_{21}$ arise despite the presence of NME uncertainty $\xi$ in (\ref{eq:chi2}). To give the readers a feeling we focus on the case of almost degenerate mass spectrum discussed in section~\ref{subsec:degenerate}. In this case $m_{0\nu\beta\beta}$ determined by nature is related to the ``observed'' one, the inferred one with use of particular NME as $m_{0\nu\beta\beta}^{\text{obs}} = \xi^{-1} m_{0\nu\beta\beta}^{\text{true}}$, where $m_{0\nu\beta\beta}^{\text{true}}$ is given by (\ref{eq:0nubb-degenerate}). Since $m_{0\nu\beta\beta}^{\text{obs}} \propto \xi^{-1} m_0$, if we are completely ignorant about $m_0$, there is no way to constrain the value of $\alpha_{21}$ even in the case of no NME uncertainty ($\xi = 1$) no matter how accurately $m_{0\nu\beta\beta}$ itself is determined.
Therefore, the first requirement for a sensitivity to the Majorana phases is to have an independent measurement of absolute neutrino mass scale other than $0\nu\beta\beta$ decay.\footnote{
We note, however, that this is not always true for the case of non-degenerate neutrino mass spectrum, in particular in region $m_0 \ll 0.1$ eV. In certain cases it would be possible to constrain the CP phase only using the information from $0\nu\beta\beta$ decay experiment.
} 
The second requirement is that the uncertainty of NME is not too large. 
We note that 
the variation of $m_{0\nu\beta\beta}^{\text{true}}$ is as large as 
a factor of $\simeq$ 2.6 as $\alpha_{21}$ is varied from 0 to $2\pi$.
(see section~\ref{subsec:degenerate}). 
Then, $\alpha_{21}$ can be constrained only if 
the NME uncertainty cannot compensate the variation.

\begin{figure}[!h]
\hglue 0.95cm
\includegraphics[width=0.90\textwidth]{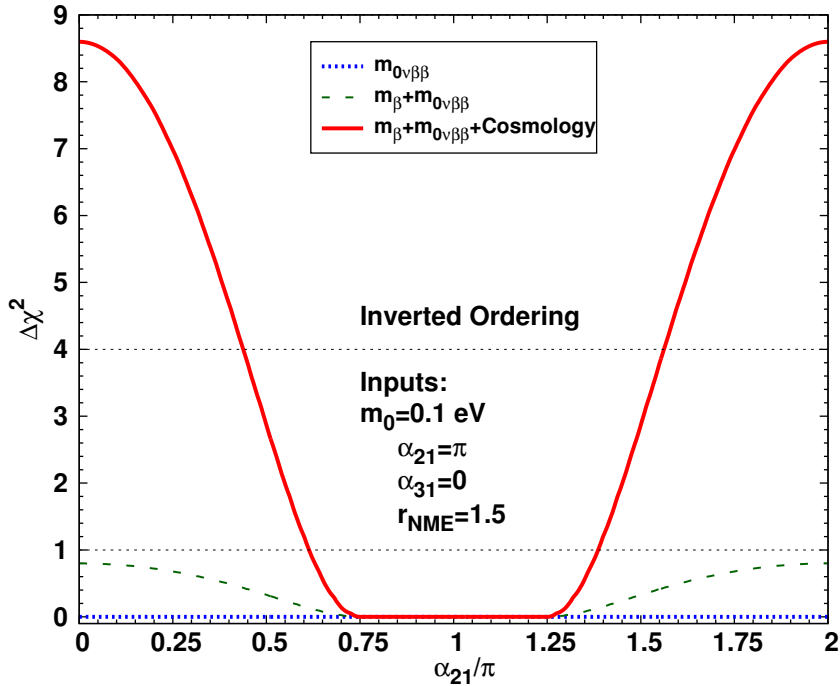}
\vglue -1.0cm
\caption{
$\Delta \chi^2$ is plotted as a function of the fitted value 
of $\alpha_{21}$ for the case of the inverted mass ordering. 
The true values of the parameters are taken as $m_0 = 0.1$ eV, 
$\alpha_{21} = \pi$, and $\alpha_{31} = 0$. 
The three $\Delta \chi^2$ curves are presented which correspond to three
 combinations of the data used in the analysis, only $0\nu\beta\beta$
 decay (dotted blue line), $0\nu\beta\beta$ + $\beta$ decays (dashed
 green curve) and all combined, $0\nu\beta\beta$ + $\beta$ decays +
 cosmology (solid red curve). 
In the fit the remaining parameters, $m_0$  and $\alpha_{31}$, 
are marginalized. 
}
\label{fig:Delta_chi2-alpha21}
\end{figure}

Let us now examine in more detail how non-vanishing $\Delta \chi^2$
develops to produce the constraint on $\alpha_{21}$. In
Fig.~\ref{fig:Delta_chi2-alpha21} the contributions to $\Delta \chi^2$
from the three types of experiments are plotted as a function of
$\alpha_{21}$ for the case of inverted mass ordering. The dotted blue,
dashed green, and the solid red curves indicate, in order, $\Delta
\chi^2$ obtained when only $0\nu\beta\beta$ decay experiment is
considered, when the $\beta$ decay data is added, and all the three
experiments are considered (i.e., the cosmological measurement of
$\Sigma$ is also added). 
As it stands, the single beta and $0\nu\beta\beta$ decay experiments 
cannot constrain the value of $\alpha_{21}$ at CL higher than $1
\sigma$. We confirmed that this feature is true even if there is no NME
uncertainty. But, once the cosmological observation of $\Sigma$ is
added, 
$\Delta \chi^2$ jumps in region outside $0.6 \pi -1.4 \pi$ of $\alpha_{21}$, as shown by the solid red curve in Fig.~\ref{fig:Delta_chi2-alpha21}. 
We, however, stress that the relative strength of the experiments in constraining $\alpha_{21}$ heavily relies on our error estimates in (\ref{reference-error}). 
If we repeat the similar exercise as in
Fig.~\ref{fig:Delta_chi2-alpha21} with the normal mass ordering, we
observe that the sensitivity to the Majorana phase is lower. The shape
of $\Delta \chi^2$ is very similar to that of the inverted ordering, but
the height is about $2/3$ of 
the $\Delta \chi^2$ curves in Fig.~\ref{fig:Delta_chi2-alpha21}.\footnote{
It is because, for a given value of $m_0$, the first and second terms which contain $m_1$ and $m_2$, respectively, in (\ref{eq:0nubb-1}) are always larger in the case of inverted ordering than the ones in the normal one, as we can see from (\ref{eq:mass-normal}) and (\ref{eq:mass-inverted}). Assuming the difference between $\vert \Delta m^2_{32} \vert $ in both mass ordering is small, as is the case in (\ref{eq:best-fit2}), it makes $\Delta \chi^2$ for the inverted mass ordering larger, thereby making the inverted mass ordering case more sensitive to the change of $\alpha_{21}$. 
}

Therefore, in region $m_0 \gsim 0.1$ eV, $0\nu\beta\beta$ decay
experiments can produce nontrivial constraint on $\alpha_{21}$ by
circumventing the NME uncertainty, but only when it is combined with
precision measurement of absolute mass scale of neutrinos. We note that
at the bottom of $\Delta \chi^2$ 
in the region where $\alpha_{21} = 0.75 \pi -1.25 \pi$, 
$\chi^2_{\text{min}}$ is very flat because of the NME uncertainty. It means that it is impossible to pin down, in the presence of the NME uncertainty, the value of the Majorana phase $\alpha_{21}$ at the bottom no matter how accurately all the measurements are done.

In the case of inverted mass ordering with hierarchical spectrum, $m_{0\nu\beta\beta}$ is given by (\ref{IH-hierarchical}), and the role of $m_0$ in the above discussion is played by $\sqrt{\Delta m^2_{\text{atm}} }$. Therefore, an accurate measurement of $\Delta m^2_{\text{atm}}$, in principle, replaces the cosmological measurement. However, $m_{0\nu\beta\beta}$ is small, $m_{0\nu\beta\beta} \sim 0.05$ eV, in this region, and the error taken in (\ref{reference-error}) is relatively large. Then, we do not expect significant sensitivity to the Majorana phase. This last statement applies also to the normal mass ordering with hierarchical spectrum. These features as well as the ones discussed above for the degenerate mass spectrum will be demonstrated in the next section.

\section{Analysis results: Sensitivity to the Majorana phase}
\label{sec:results-exclusion-fraction}

In this section, we show the results of our analysis on sensitivity to 
the Majorana phase $\alpha_{21}$ by using the CP exclusion fraction 
$f_{\text{CPX}}$ \cite{Machado:2013kya}. 
It is defined as the fraction of values of $\alpha_{21}$ $\in [0,2\pi]$ which can be excluded by the experiments at a given confidence level for each input point of the parameter space $(m_0, \alpha_{21})$ and a given NME uncertainty. It is a global measure for CP sensitivity and may be particularly useful in the initial stage in which the sensitivity to the CP phase may be limited. For more about $f_{\rm CPX}$ see Ref.~\cite{Machado:2013kya}. 

\subsection{CP exclusion fraction: the case of reference errors}
\label{sec:reference}

\begin{figure}[!h]
\begin{center}
\hglue -7mm
\includegraphics[width=0.99\textwidth]{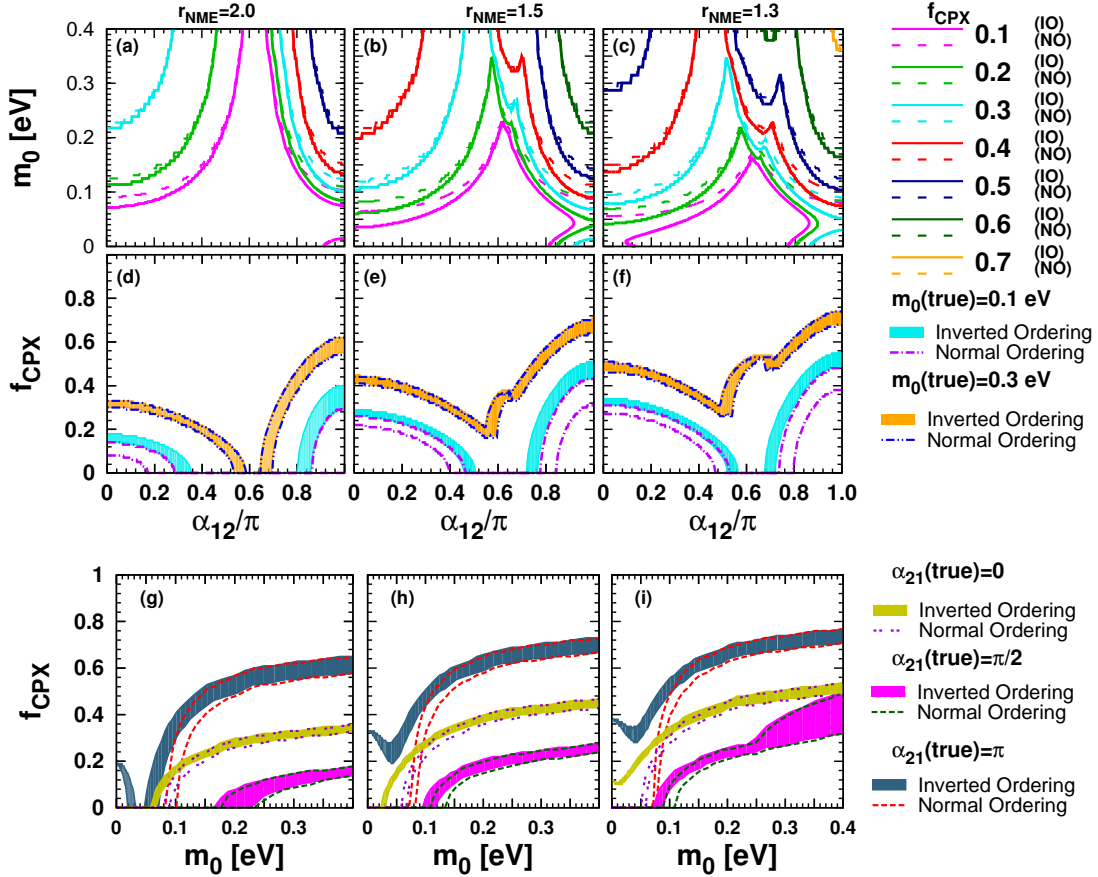}
\end{center}
\vglue -3mm
\caption{
In the first row [panels (a), (b), and (c)]: shown are the iso-contours of the CP exclusion fraction determined at $2\sigma$ (95.45\%) CL (1 DOF) projected into the plane of the true values of $\alpha_{21}/\pi$ and the lightest neutrino mass $m_0$. 
In the second row [panels (d), (e), and (f)]: $f_{\text{CPX}}$ is plotted as a function of the true values of $\alpha_{21}$ at $m_0=0.1$ and 0.3 eV, which is nothing but slices of the iso-contours of $f_{\text{CPX}}$ at the values of $m_0$. 
In the third row [panels (g), (h), and (i)]: $f_{\text{CPX}}$ as a function of $m_0$ at $\alpha_{21} = 0, \pi/2$, and $\pi$. 
In each row the left, middle and the right panels are for uncertainties of the NME $r_{\text{\tiny NME}}=$ 2.0, 1.5, and 1.3, respectively.  
The case for the inverted and normal mass orderings are shown, respectively, by the solid and dashed curves from 0.1 to 0.7 with the step size of 0.1 in the first row, and by using the colored band and dashed (dotted or dash-dotted) lines, respectively, in the remaining rows. 
}
\label{f_CPX-2s-2-1.3-0.05eV}
\end{figure}

We show in the three rows in Fig.~\ref{f_CPX-2s-2-1.3-0.05eV} 
the results obtained for the CP exclusion fraction $f_{\text{CPX}}$, 
which indicates that some sensitivities to the Majorana phase do indeed exist.
(1) First row: shown are the iso-contours of the CP exclusion fraction
determined at $2\sigma$ (95.45\%) CL (1 DOF) in the plane spanned by 
the true values of $\alpha_{21}$ and the lightest neutrino mass $m_0$. 
Both $\alpha_{21}$ and $m_0$ are varied in the fit.
(2) Second row: $f_{\text{CPX}}$ is plotted as a function of the 
true values of $\alpha_{21}$ at $m_0=0.1$ and 0.3 eV 
where input value of $m_0$ is fixed but it is varied in the fit. 
Roughly speaking, it is nothing but the cross section of 
the iso-contours of $f_{\text{CPX}}$ at the values of $m_0$. 
(3) Third row: $f_{\text{CPX}}$ as a function of true values of $m_0$ at 
$\alpha_{21} = 0, \pi/2$, and $\pi$. 
Similar to the case of (2), input value of $\alpha_{21}$ 
is fixed but it is varied in the fit.
\footnote{
To obtain $f_{\text{CPX}}$ plots in the second and third rows 
marginalization over the fitted parameters is carried out, 
which produces a finite width of the $f_{\text{CPX}}$ curves.
}
In each row the left, middle and the right panels are 
for uncertainties of the NME, $r_{\text{\tiny NME}}=$ 2.0, 1.5, and 1.3, respectively.  
The second and third rows are provided for ease of understanding the complex structure of the iso-contours of $f_{\text{CPX}}$ given in the first row. 
Due to the symmetry (\ref{eq:symmetry}), the iso-contours of $f_{\text{CPX}}$ are symmetric under reflection around $\alpha_{21}=\pi$ after marginalizing over $\alpha_{31}$. Hence, we show the results only for the range of $0 \leq \alpha_{21} \leq \pi$. Notice that larger the values of $f_{\text{CPX}}$, higher the sensitivity to $\alpha_{21}$ because larger fraction of the phase space can be excluded. 

In the first row, the iso-contours of $f_{\text{CPX}}$ for the inverted and the normal mass orderings are depicted, respectively, by the solid and the dashed lines, from 0.1 to 0.7 (with the step size of 0.1), using different colors as indicated in the legend. In the second and third rows the case of inverted (normal) mass ordering is depicted by using colored band (dotted, dashed, or dash-dotted lines). We note that for each point on the input parameter space, all the parameters, $m_0$, $\alpha_{21}$, $\alpha_{31}$ and $\xi$ were varied in doing the fit. 
In Fig.~\ref{f_CPX-2s-2-1.3-0.05eV} the true value of $\alpha_{31}$ is taken to be $0$, and the similar contour plot with $\alpha_{31} = \pi$ (not shown here, see \cite{Minakata:2014jba}) indicates that the sensitivity is slightly higher but not much.\footnote{
We have examined the similar $f_{\text{CPX}}$ plots for several input values of $\alpha_{31}$, $0$, ${\pi}/{4}$, ${\pi}/{2}$, and $\pi$. The exercise revealed that for $\alpha_{21} \gsim 0.6\pi$, the best and the worst sensitivities are obtained at $\alpha_{31}=\pi$ and $\alpha_{31}=0$, respectively, the cases used in Fig.~\ref{f_CPX-2s-2-1.3-0.05eV}. For $\alpha_{21} \lsim 0.6\pi$, this behaviour become opposite, leading to the worst and best sensitivity at $\alpha_{31}=\pi$ and $\alpha_{31}=0$, respectively, 
} 
%

%

As can be seen most clearly in the second row of
Fig.~\ref{f_CPX-2s-2-1.3-0.05eV}, the sensitivity to $\alpha_{21}$ is
highest in region around $\alpha_{21} \simeq \pi$ and next highest in
region of $\alpha_{21}$ near 0. The worst sensitivity is obtained at
around $\alpha_{21} \sim {2\pi}/{3}$, in agreement with the qualitative
discussion given in section~\ref{sec:analytic-estimation}. Yet, it may
be interesting to note that the sensitivity improves at around the worst
sensitive region if the NME errors are controlled better, 
as can be seen in the middle and right panels (e) and (f) in Fig.~\ref{f_CPX-2s-2-1.3-0.05eV}. As expected the
difference between the normal and the inverted mass orderings are small
in the degenerate regime, $m_0 \gsim 0.1$ eV where 
the colored bands (inverted) and dashed double dotted line (normal) in the second and third rows, overlap rather well. A better sensitivity to the phase $\alpha_{21}$ for the inverted than the normal ordering case can be observed around region $m_0 \sim 0.05$ eV where the sensitivity is however rather low.

It is important to notice that the sensitivity to $\alpha_{21}$ strongly depends upon $r_{\text{\tiny NME}}$. In particular the improvement of the sensitivity from $r_{\text{\tiny NME}}=2.0$ to $r_{\text{\tiny NME}}=1.5$ is remarkable. At $m_0 = 0.15$ eV, for example, while $f_{\text{CPX}} \ge$ 0.3 region spans only $\simeq 0.2$ of $\alpha_{21}$ space for $r_{\text{\tiny NME}}=2.0$, it jumps to $\simeq 50\%$ coverage for $r_{\text{\tiny NME}}=1.5$ for both cases of $\alpha_{31} = 0$ and $\pi$. 
It is also notable that in the inverted mass ordering case the sensitivity region reaches to the range $0~\text{eV} \leq m_0 \lsim 0.05~\text{eV}$, though only up to $f_{\text{CPX}} \simeq$ $0.1- 0.2$.\footnote{
We note here that we took a fixed error (0.01 eV) for $m_{0\nu\beta\beta}$ based on the discussion given in Appendix~\ref{sec:double-beta-sensitivity}. If a percentage error for $m_{0\nu\beta\beta}$ is to be assumed, the sensitivity to $\alpha_{21}$ in region $m_0 \lsim 0.05$ eV would be greatly enhanced. 
}
The tendency is more and more visible for smaller values of $r_{\text{\tiny NME}}$. Therefore, reducing the NME uncertainty is of crucial importance to have severer constrains on (or to observe) the Majorana phase.

\begin{figure*}[!h]
\hglue 5mm
\includegraphics[width=0.88\textwidth]{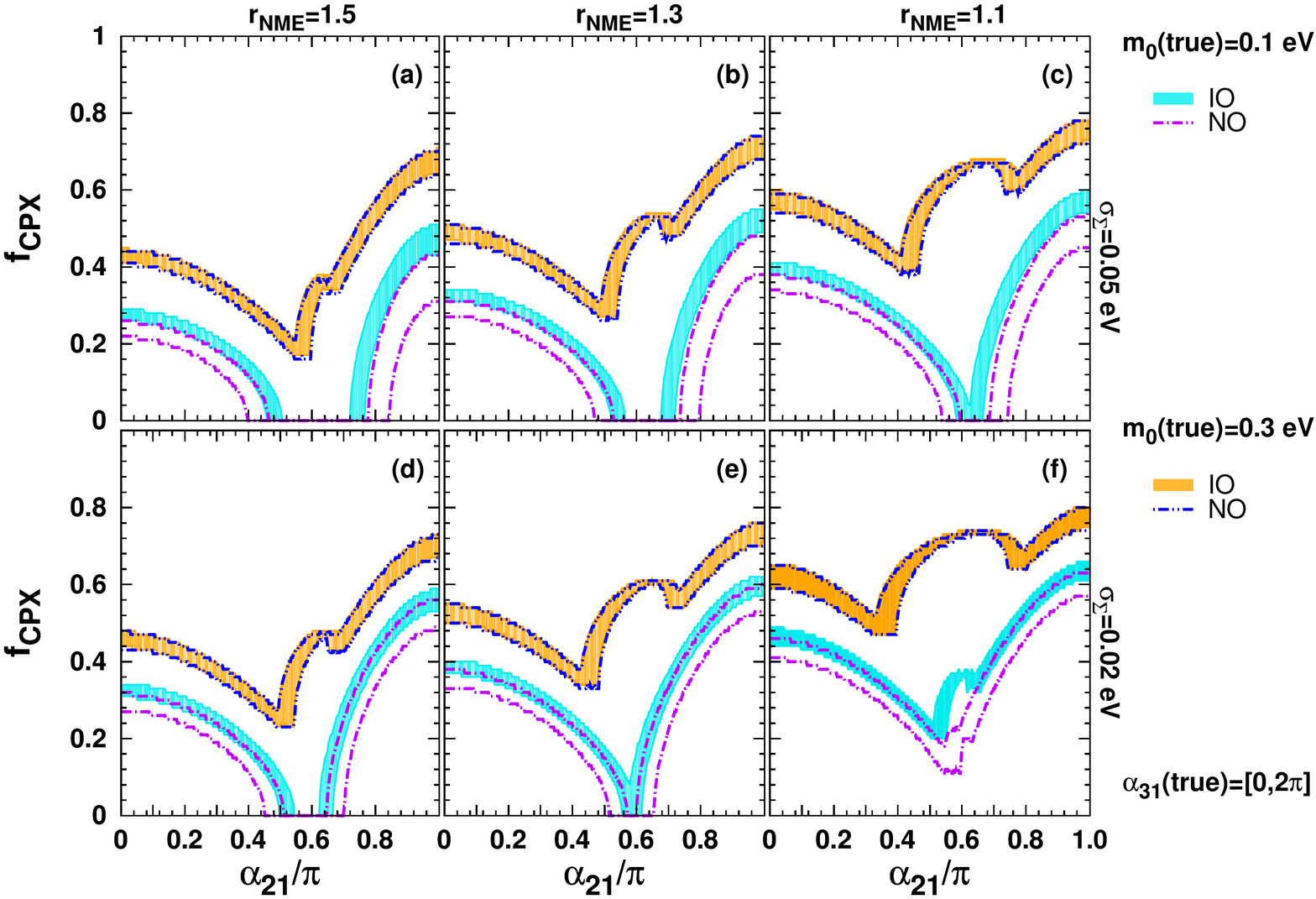}
\vglue -15mm
\hglue 5mm
\includegraphics[width=0.88\textwidth]{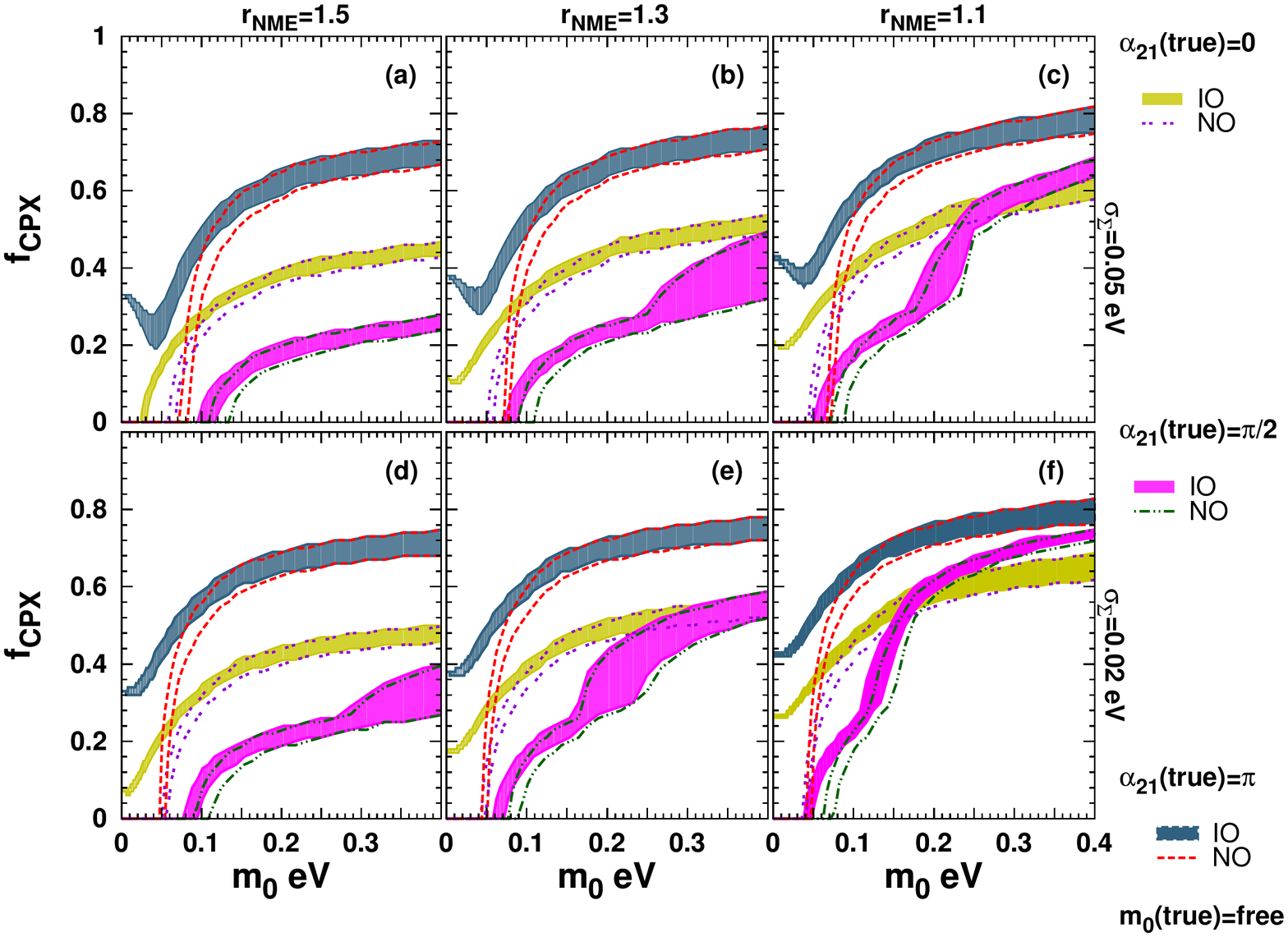}
\vglue -0.9cm
\caption{
In the upper and the lower clusters of six panels, shown are $f_{\text{CPX}}$ as a function of the true value of $\alpha_{21}$ at $m_0=0.1$ and 0.3 eV, and $f_{\text{CPX}}$ as a function of the true value of $m_0$ at $\alpha_{21}=0, \pi/2$, and $\pi$, respectively. In both clusters the true value of ($\sigma_\Sigma$, $r_{\text{\tiny NME}}$) used are: 
(a) (0.05 eV, 1.5),  (b) (0.05 eV, 1.3),  (c) (0.05 eV, 1.1), (d) (0.02 eV, 1.5),  (e) (0.02 eV, 1.3), and (f) (0.02 eV, 1.1). 
}
\label{f_CPX-optimistic-error}
\end{figure*}

\subsection{CP exclusion fraction: the case of more optimistic errors}
\label{sec:optimistic}

Though it is encouraging to see some sensitivities to the Majorana phase
thanks to precision measurement of $\Sigma$ and $0\nu\beta\beta$ decay
rate, we must remark that the sensitivity is still at a relatively low
level. For example, in the case of $m_0 \simeq 0.1$ eV, which is
preferred by some cosmological analyses, the sensitivity to
$\alpha_{21}$ is up to the level of $f_{\text{CPX}}$ $\simeq 0.1-0.3$
even in the most optimistic case of $r_{\text{\tiny NME}} = 1.3$ except
for the region close to $\alpha_{21} = \pi$ (see the second row of
Fig.~\ref{f_CPX-2s-2-1.3-0.05eV}). Moreover, there always exists 
a region of $\alpha_{21}$ in which $f_{\text{CPX}}$ vanishes 
at around the worst sensitive region, and the region is 
quite wide for $r_{\text{\tiny NME}} = 2.0$. 

Therefore, we examine below two possibilities toward further improvement of the sensitivity to $\alpha_{21}$: 
(1) Higher precision measurement of neutrino masses in cosmology as good as $\sigma_\Sigma=0.02$ eV, and 
(2) Revolutionary new technology of computing the NME of $0\nu\beta\beta$ decay which could lead to uncertainty of the level $r_{\text{\tiny NME}} = 1.1$. 
As mentioned before, once the positive signal of $0\nu\beta\beta$ will be observed by using different isotopes, we hope that theoretical NME calculations can be ``calibrated'' to some extent using the real data. (Nonetheless, we tentatively assume that 10\% level uncertainty is unavoidable.) It is discussed that the former is within reach in the light of future cosmological observation as mentioned in section~\ref{subsec:sum-mass} \cite{Carbone:2010ik,Kitching:2008dp,Amendola:2012ys}. On the other hand, it remains to be seen how (2) can be realized. 

In Fig.~\ref{f_CPX-optimistic-error} we present the results of $f_{\text{CPX}}$ for these more optimistic experimental and theoretical uncertainties. In the upper and the lower clusters of six panels of Fig.~\ref{f_CPX-optimistic-error}, we show $f_{\text{CPX}}$ as a function of the true value of $\alpha_{21}$ at $m_0=0.1$ and 0.3 eV, and $f_{\text{CPX}}$ as a function of the true value of $m_0$ at $\alpha_{21}=0, \pi/2$, and $\pi$, respectively. In each cluster of panels the upper (lower) three panels are for cases with $\sigma_{\Sigma}$ (the error of $\Sigma$) of 0.05 eV (0.02 eV). The panels (a) and (b) both in the upper and the lower clusters in Fig.~\ref{f_CPX-optimistic-error} are identical to the panels (e) and (f) in Fig.~\ref{f_CPX-2s-2-1.3-0.05eV}; They are duplicated to make comparison with the corresponding cases with the smaller uncertainties in the lower panels easier. 

We observe that, globally, the sensitivity increases from left to right
($r_{\text{\tiny NME}} = 1.5, 1.3, 1.1$), and from the upper to lower
panels ($\sigma_{\Sigma}=0.05, 0.02$ eV). However, the improvement in
the sensitivity to the phase is relatively modest despite the highly
nontrivial requirement on the uncertainties of $r_{\text{\tiny NME}}$ or
$\sigma_{\Sigma}$. Most notable exception may be that $f_{\text{CPX}}
\neq 0$ at $m_0=0.1$ eV is realized for the first time within our
analysis, which however, occurs for an extreme case of 
$r_{\text{\tiny NME}} = 1.1$ and $\sigma_{\Sigma} = 0.02$ eV. 
There exists significant variation in the sensitivity to the phase as 
the true value of $\alpha_{21}$ is varied (lower cluster of panels). 
The similar dependence is also visible for varying $m_0$
(upper cluster of panels). 
A good news is that at $m_0=0.1$ eV, the average value of 
$f_{\text{CPX}}$ reaches $\simeq 0.3$ for 
the case $\sigma_{\Sigma}=0.02$ eV as seen in the panel (e) in 
the upper cluster in Fig.~\ref{f_CPX-optimistic-error}. 
In all the panels in the lower cluster the difference of 
$f_{\text{CPX}}$ curves between the inverted and 
the normal mass orderings are quite visible in region at $m_0 \lsim 0.1$ eV.

One may say that the sensitivity to $\alpha_{21}$ is low, i.e.,
excluding only 30\% of its phase space at $2 \sigma$ CL, which may be
insufficient to foresee the measurement of the Majorana phase. 
Nonetheless, we hope that the results we report in this paper would 
imply a nontrivial step to obtain perspectives for measuring 
the Majorana phase. 
Clearly, imaginative discussions on how to improve 
the sensitivity to the phase are called for.

\section{Conclusions and discussion}
\label{sec:conclusions}

Assuming the mass mechanism of $0\nu\beta\beta$ decay, we have studied in this paper how and to what extent the Majorana phases can be constrained in the light of future accurate measurements of $0\nu\beta\beta$ decay rate and an independent precision measurement of neutrino's absolute mass scale. In our estimates of the errors described in Sec.~\ref{sec:exp-errors} cosmological observation of the sum of neutrino masses $\Sigma$ plays the latter role. We have demonstrated that even in the situation that double beta decay experiment alone cannot say anything about the Majorana phase, it is possible to obtain highly non-trivial constraints on the Majorana phase, if accurate measurement of $\Sigma$ is added. Indeed, it is unlikely that $0\nu\beta\beta$ decay experiment by itself can place any useful constraints on the Majorana phase within our assumption of the experimental errors. Notice, however, that it is not because they suffer from the NME uncertainty (which of course makes the situation worse), but because they lack independent key information of the absolute neutrino mass scale in itself (see Sec.~\ref{how-constrained}).
The remarkable feature of our results is that the synergy between cosmology and double beta decay experiments in constraining the Majorana phase is quite visible.

The general tendency of the sensitivity to the Majorana phases are as follows: 
\begin{itemize}

\item 
Dependence of $m_{0\nu\beta\beta}$ on $\alpha_{31}$, one of the two Majorana phases, is very weak due to the suppression by $s^2_{13}$. Hence, little sensitivity can be expected for $\alpha_{31}$. The sensitivity to $\alpha_{21}$ depend on $\alpha_{31}$, but only very weakly. 

\item 
The regions of the best and the worst sensitivities to $\alpha_{21}$ are located at around the true values $\alpha_{21} \simeq 0$ or $\pi$, and at $\alpha_{21} \simeq {2\pi}/{3}$, respectively, in agreement with our analytic estimate in Sec.~\ref{subsec:degenerate}.

\item 
For both mass orderings and within any uncertainties of the NME used, the better sensitivity to the CP phase $\alpha_{21}$ is obtained for larger value of $m_0$, apart from minor exception of region $m_0 \lsim 0.05$ eV which is mostly in the horizontal branch in Fig.~\ref{fig:m0nbb-m0} for the inverted mass ordering. 

\end{itemize}

We took as the reference set up of the uncertainties that the effective masses observed in $0\nu\beta\beta$ and $\beta$ decay experiments can be measured with the errors of 0.01 eV and 0.06 eV, respectively, whereas the sum of neutrino masses $\Sigma$ can be determined with the uncertainty of 0.05 eV by cosmological observation. We have assumed the uncertainty factor $r_{\text{\tiny NME}}$ (defined as the ratio of the maximum and minimum values of the theoretically expected NME values) of a range $1.3 - 2.0$. We have treated it in a different way from these experimental errors, in similar way as the treatment of flux normalization uncertainty. 
To show the dependence of these assumed errors on sensitivity to Majorana phase, we have also examined the case with more optimistic errors, 0.02 eV for the error of $\Sigma$ and $r_{\text{\tiny NME}} =1.1$ for the NME uncertainty.

To display the sensitivities to the Majorana phase globally we have used the new measures, the CP exclusion fraction, $f_{\rm CPX}$, a fraction of the CP phase space that can be excluded for a given set of input parameters. $f_{\rm CPX}$ is a useful global measure, as the CP fraction widely used in the sensitivity studies for the long-baseline neutrino oscillation experiments is, but in a different way. We believe that $f_{\rm CPX}$ is better suited to reveal (relatively poor) sensitivities in the early era of the search for the CP phase effect, as discussed in \cite{Machado:2013kya}. Since measuring the Majorana phase is very challenging in any means, we believe that even the partial exclusion of its possible range is quite useful. 

We have shown with our reference setup and $r_{\text{\tiny NME}} =1.5$ that $\alpha_{21}$ can be constrained by excluding $\simeq 10-40\%$ of the phase space of $\alpha_{21}$ at $2\sigma$ CL for the lowest neutrino mass of 0.1 eV for both mass orderings. Even if we use 0.02 eV for the error of $\Sigma$ the excluding fraction does not change much, and the exclusion fraction is $\simeq 10-50\%$. Fortunately, the $f_{\rm CPX}$ contours are rather stiff against change in confidence level from $2\sigma$ to $3\sigma$, indicating robust nature of sensitivity to the Majorana phase. We also should note that the sensitivity to $\alpha_{21}$ becomes significantly better when the uncertainty of the nuclear matrix element ($r_{\text{\tiny NME}}$) is reduced from the factor 2 to 1.5, independent of the assumed mass ordering and value of $\alpha_{31}$. 

It is only recently that such a discussion started to obtain real perspective because cosmology entered into the precision era, as demonstrated most notably by an epoch making precision measurement achieved by Planck. Yet, the accuracy of measurement of the sum of neutrino masses which warrants reasonably strong restriction to the Majorana phase is rather demanding, $\sigma_{\Sigma} \simeq 0.02$ eV, which probably requires the Euclid satellite as well as the next generation galaxy surveys, in addition to the better understanding of the different systematics exist in the different set of the cosmological data. Of course, it also requires accurate measurement of lifetime of $0\nu\beta\beta$ decay in a ton scale experiment with very low background, which would allow uncertainty in measurement of $m_{0\nu\beta\beta}$ as small as $\simeq 0.01$ eV.

One may argue that too much relying on cosmological observation in
determining $\Sigma$ is dangerous because, neutrinos being a minor
player in the universe, its precise determination is possible only in a
model dependent way. Though it is a valid point we must bear in mind
that future precision observation itself offers an even more stringent
test of the $\Lambda$CDM paradigm. Then, it may be conceivable that the
SM of cosmology could be eventually born out from such process, which we
assumed as a prerequisite of our analysis. It would allow us a reliable
measurement of the neutrino component in the universe and to determine
the sum of their masses with a sufficient robustness. The discussion in
this paper would serve as a prototype of the similar analysis 
that should be done after we have acquired 
the established standard cosmological model. 

\appendix

\section{Estimation of the sensitivity to $m_{0\nu\beta\beta}$ in neutrinoless double beta decay experiment}
\label{sec:double-beta-sensitivity}

Let us estimate the expected precision 
on $m_{0\nu\beta\beta}$ to be observed in 
the $0\nu\beta\beta$ decay experiment. 
See Refs.~\cite{Elliott:2002xe,Avignone:2005cs,Avignone:2007fu,GomezCadenas:2011it,Giuliani:2012zu,Cremonesi:2013bma}
for more detailed discussions on the experimental sensitivities. 
Here we ignore the theoretical uncertainty 
of the nuclear matrix element, ${\cal{M}}^{(0\nu)}$, 
which will be taken care of in a different way  as discussed in 
section~\ref{sec:analysis-method}. 

From Eq.~(\ref{eq:decay-rate})
we obtain the following expression for the effective mass, 
\begin{equation}
m_{0\nu\beta\beta} = 
\frac{m_e}{\sqrt{
T_{1/2}^{0\nu}G_{0\nu} \left|{\cal{M}}^{(0\nu)}\right|^2 }}.
\label{eq:m0nunu-NME}
\end{equation}
The expected number of $0\nu\beta\beta$ decays 
(signal) to be observed 
in the experiment,
$N_{0\nu\beta\beta}$, is given by 
\begin{equation}
N_{0\nu\beta\beta} = 
\varepsilon_{\text{det}} \frac{m_X N_A}{W_X}
\left[1-\exp\left(-\frac{t_{\text{exp}}\ln2 }{T_{1/2}^{0\nu}}
\right) \right]
\simeq 
\frac{ \varepsilon_{\text{det}} N_A m_X t_{\text{exp}}\ln2}{W_X T_{1/2}^{0\nu}}
\label{eq:N_decay}
\end{equation}
where $\varepsilon_{\text{det}}$ is the detection efficiency, 
$m_X$ and $W_X$ are respectively, the total mass and
the molecular weight of the isotope $X$ to be used in 
the double beta decay experiment, 
$N_A$ is the Avgadro's number and 
$t_{\text{exp}}$ is the exposure time, assumed
to be much smaller than $T_{1/2}^{0\nu}$. 

On the other hand, the expected number of the background 
events can be expressed as, 
\begin{equation}
N_{\text{BG}}  = b \Delta E m_X t_{\text{exp}}, 
\label{eq:BG}
\end{equation}
where $b$ is the number of background counts usually measured 
in keV$^{-1}\cdot$ kg $^{-1}\cdot$ yr$^{-1}$, $\Delta E$ is the energy 
window ($\sim$ energy resolution) given in keV around 
the $0\nu\beta\beta$ peak, 
both of which depend on the experimental set up we consider.

Roughly speaking, the expected sensitivity of the double beta 
decay experiment for the half life time is obtained when 
$N_{0\nu\beta\beta}\sim \sqrt{N_{\text{BG}}}$, which implies 
that 
\begin{equation}
T_{1/2}^{0\nu} 
\sim \frac{\varepsilon_{\text{det}} N_A m_X t_{\text{exp}} \ln 2 }
{W_X \sqrt{b \Delta E m_X t_{\text{exp} }}
}
= \frac{\varepsilon_{\text{det}} N_A\ln 2} {W_X}
\sqrt{  \frac{ m_X t_{\text{exp}}}{b \Delta E} 
}.
\end{equation}

This can be translated in terms of the minimum 
possible observable value of  
$m_{0\nu\beta\beta}^{\text{min}}$ as~\cite{Moe:1991ku}, 
\begin{equation}
m_{0\nu\beta\beta}^{\text{min}}
\sim 
\frac{m_e}{
\sqrt{G_{0\nu} \left|{\cal{M}}^{(0\nu)}\right|^2 
\ln 2 }}
\left[\frac{W_X}{\varepsilon_{\text{det}} N_A} \right] ^{\frac{1}{2}}
\left[ \frac{b \Delta E} {m_X t_{\text{exp}}}\right] ^{\frac{1}{4}}.
\end{equation}
If we consider, for example, the isotopes of $^{76}$Ge and 
$^{136}$Xe, 
by using the typical values for $G_{0\nu}$ and 
${\cal{M}}^{(0\nu)}$ found, e.g.,  
in Ref.~\cite{Cremonesi:2013bma}, 
and typical background rate and energy resolutions, 
for $^{76}$Ge (by using 
$G_{0\nu} = 2.36 \times 10^{-15}$ yr $^{-1}$~\cite{Kotila:2012zza})
we obtain, 
\begin{equation}
\hskip -0.5cm
m_{0\nu\beta\beta}^{\text{min}}
\sim 
0.12\  
\left[ \frac {5.0}{ {\cal{M}}^{(0\nu)} } \right] 
\left[ \frac {b}
{ 0.01 \ \text{keV}^{-1} 
\cdot\text{kg}^{-1}\cdot\text{yr}^{-1} } \right] ^{\frac{1}{4}}
\left[ \frac {\Delta E}{ 3.5\ \text{keV} } \right] ^{\frac{1}{4}}
\left[ \frac{100\ \text{kg}\cdot{\text{yr}}}
{\varepsilon_{\text{det}}^2\cdot 
m_{\text{Ge}}\cdot t_{\text{exp}}}\right]^{\frac{1}{4}}\ \text{eV},
\label{eq:m0bb_min_Ge}
\end{equation}
whereas for $^{136}$Xe 
(by using 
$G_{0\nu} = 14.58 \times 10^{-15}$ yr$^{-1}$~\cite{Kotila:2012zza}) 
we obtain, 
%
\begin{equation}
\hskip -0.5cm
m_{0\nu\beta\beta}^{\text{min}}
\sim 0.24\  
\left[ \frac {3.0}{ {\cal{M}}^{(0\nu)} } \right] 
\left[ \frac {b}
{ 0.01 \ \text{keV}^{-1}\cdot\text{kg}^{-1}\cdot\text{yr}^{-1} } \right] ^{\frac{1}{4}}
\left[ \frac {\Delta E}{ 100\ \text{keV} } \right] ^{\frac{1}{4}}
\left[ \frac{100\ \text{kg}\cdot{\text{yr}}}
{\varepsilon_{\text{det}}^2\cdot m_{\text{Xe}}\cdot 
t_{\text{exp}}}\right]^{\frac{1}{4}}
\ \text{eV},
\label{eq:m0bb_min_Xe}
\end{equation}
where $m_{\text{Ge}}$ ($m_{\text{Xe}}$) is the total mass of 
the isotope of $^{76}$Ge ($^{136}$Xe) to be used in
the double beta decay experiments.
If $0\nu\beta\beta$ decay will be actually observed, 
at first approximation, we assume that these values 
could be roughly corresponds
to the uncertainty on the measurement of 
$m_{0\nu\beta\beta}$, 
or $\sigma_{0\nu\beta\beta}$.

As we can see, as the sensitivity improves only 
as the one of the fourth power of the size of the experiment, 
background and energy resolution, 
it is seems not so easy to improve the sensitivity 
and looks quite difficult to reach the level of 
$\sigma_{0\nu\beta\beta}$ $\sim O(0.01)$ eV, 
even if we consider $\sim$ 1 ton scale experiment. 
Therefore, it would be necessary to realize 
the background free or very low background
experiment to achieve 
such a level (see below). 

Next let us consider the case that the background rate $b$ is so low
that  $N_{BG}$ in Eq.~(\ref{eq:BG}) is negligible, 
or namely, {\it zero background} experiment~\cite{Cremonesi:2013bma}. 
In this case, from the number of observed events, 
$N_{0\nu\beta\beta}$, 
the half life time can be estimated as 
\begin{equation}
T_{1/2}^{0\nu} = \frac{\varepsilon_{\text{det}} n_X t_{\text{exp}} \ln 2 }
{ N_{0\nu\beta\beta} }, 
\label{eq:lifetime}
\end{equation}
and its uncertainty is estimated as
\begin{equation}
\delta (T_{1/2}^{0\nu}) 
\sim 
 T_{1/2}^{0\nu} 
\frac{ \delta( N_{0\nu\beta\beta} )}{ N_{0\nu\beta\beta}  }
\sim 
 T_{1/2}^{0\nu} 
\frac{ 1}{ \sqrt{ N_{0\nu\beta\beta} } }.
\label{eq:delta-lifetime}
\end{equation}
Then from Eqs.~(\ref{eq:m0nunu-NME}) and (\ref{eq:delta-lifetime})
we obtain 
\begin{eqnarray}
\sigma_{0\nu\beta\beta}\equiv 
\delta(m_{0\nu\beta\beta}) 
& \sim &
\frac{1}{2}\ 
m_{0\nu\beta\beta}^{(0)}\ 
\frac{\delta(T_{1/2}^{0\nu}) }{T_{1/2}^{0\nu}}
\sim 
\frac{1}{2}\ 
m_{0\nu\beta\beta}^{(0)}\ 
\frac{ 1}{ \sqrt{ N_{0\nu\beta\beta} } } \nonumber \\
&\sim &
 \frac{m_e}{2
\sqrt{G_{0\nu} 
\left|{\cal{M}}^{(0\nu)}\right|^2 \varepsilon_{\text{det}}
(m_XN_A/W_X) t_{\text{exp}} \ln 2 }
}.
\label{eq:delta-mass2}
\end{eqnarray}
We note that the result does not depend on the life time, 
and the
uncertainty can be arbitrarily small
as we increase the product of the source mass and exposure time of 
the experiment by considering only the statistical error, 
until the point that the background could no longer be neglected. 

By using the same numbers for 
$G_{0\nu}$ and ${\cal{M}}^{(0\nu)}$ we used to obtain 
$m_{0\nu\beta\beta}^{\text{min}}$ in 
Eqs.~(\ref{eq:m0bb_min_Ge}) and (\ref{eq:m0bb_min_Xe}) and ), 
for $^{76}$Ge, we obtain, 
\begin{equation}
\sigma_{0\nu\beta\beta}
\sim
0.06\  
\left[ \frac{100\ \text{kg}\cdot{\text{yr}}}
{\varepsilon_{\text{det}}\cdot m_{\text{Ge}}
\cdot t_{\text{exp}}}\right]^{\frac{1}{2}}
\left[ \frac {5.0}{ {\cal{M}}^{(0\nu)} } \right] 
\ \ \text{eV},
\end{equation}
whereas for $^{136}$Xe, we obtain, 
\begin{equation}
\sigma_{0\nu\beta\beta} \sim
0.04\  
\left[ \frac{100\ \text{kg}\cdot{\text{yr}}}
{\varepsilon_{\text{det}}\cdot m_{\text{Xe}}
\cdot t_{\text{exp}}}\right]^{\frac{1}{2}}
\left[ \frac {3.0}{ {\cal{M}}^{(0\nu)} } \right] 
\ \ \text{eV}.
\end{equation}
We expect that these values should coincide roughly 
with the minimum possible observable values (or sensitivities)
of the experiments, or 
$\sigma_{0\nu\beta\beta}
\sim m_{0\nu\beta\beta}^{\text{min}}$. 
Therefore, as long as the background is neglected, 
by considering the $\sim$ 1 ton size experiment, 
it seems possible to reach the level of 
$\sigma_{0\nu\beta\beta} \sim O(0.01)$ eV. 
In this work, for definiteness and simplicity, 
we assume that $m_{0\nu\beta\beta}$ 
can be determined with an accuracy of 0.01 eV
for a given reference value of the NME. 
See the section \ref{sec:analysis-procedure}
how to take into account the NME uncertainty. 
We note, however, that in reality, the fate of the background would not be so simple to allow analytic treatment as ours. Therefore, most probably, one needs to perform detailed numerical simulations to reliably estimate the experimental uncertainties in measurement of the double beta decay lifetime.

\section*{Acknowledgment}

The authors thank Kunio Inoue for informative correspondences on 
neutrinoless double beta decay experiment and 
Shun Saito on the cosmological determination of neutrino masses. 
H.M. is grateful to CNPq for support for his visit to the Departamento 
de F\'{\i}sica, Pontif{\'\i}cia Universidade Cat{\'o}lica do Rio de Janeiro. 
He thanks Universidade de S\~ao Paulo for the great opportunity of
stay as Pesquisador Visitante Internacional. He is also partially
supported by KAKENHI received through Tokyo Metropolitan University, 
Grant-in-Aid for Scientific Research No. 23540315, 
Japan Society for the Promotion of Science.
H.N. thanks the hospitality of Osamu Yasuda and Omar Miranda, 
respectively, at the Department of Physics of Tokyo Metropolitan 
University and of CINVESTAV-IPN where 
the final part of this manuscript was done. 
This work was supported by 
Funda\c{c}\~ao de Amparo \`a Pesquisa do Estado do 
Rio de Janeiro (FAPERJ) and Conselho Nacional de 
Ci\^encia e Tecnologia (CNPq).

\vglue 
0.5cm
\begin{center}
{\bf Note Added}
\end{center}
After the first version of this paper \cite{Minakata:2014jba} was submitted to arXiv, a preprint~\cite{Dodelson:2014tga} appeared which addresses the constraint on the Majorana phase with the same framework of combining cosmological observations with $0\nu\beta\beta$ decay experiments. It appears that a better sensitivity to the phase reported by them comes from the optimistic choice of uncertainties in $0\nu\beta\beta$ decay measurement and in the NME, $\sigma_{0\nu\beta\beta} = 0.01$ eV and $r_{\text{\tiny NME}}=1$, 
if translated to our language.



\begin{thebibliography}{99}

\bibitem{MNS}
Z.~Maki, M.~Nakagawa and S.~Sakata, 
Prog. Theor. Phys. {\bf 28}, 870 (1962).



\bibitem{Kajita:2012vc} 
  T.~Kajita,
  Adv.\ High Energy Phys.\  {\bf 2012}, 504715 (2012).

\bibitem{McDonald:2004dd} 
  A.~B.~McDonald,
  New J.\ Phys.\  {\bf 6}, 121 (2004)
  [astro-ph/0406253].

\bibitem{Inoue:2004wv} 
  K.~Inoue,
  New J.\ Phys.\  {\bf 6}, 147 (2004).

\bibitem{KM} 
M.~Kobayashi and T.~Maskawa,
Prog. Theor. Phys. {\bf 49}, 652 (1973).

\bibitem{Schechter:1980gr}
J.~Schechter and J.~W.~F.~Valle,
Phys.\ Rev.\ D {\bf 22}, 2227 (1980).

\bibitem{Bilenky:1980cx}
  S.~M.~Bilenky, J.~Hosek and S.~T.~Petcov,
Phys.\ Lett.\ B {\bf 94}, 495 (1980).

\bibitem{Doi:1980yb}
M.~Doi, T.~Kotani, H.~Nishiura, K.~Okuda and E.~Takasugi,
Phys.\ Lett.\ B {\bf 102}, 323 (1981).


\bibitem {leptogenesis}
M.~Fukugita and T.~Yanagida,
Phys.\ Lett.\ {\bf B  174} 45, (1986).

\bibitem{Minakata:1996vs}
  H.~Minakata and O.~Yasuda,
Phys.\ Rev.\ D {\bf 56}, 1692 (1997) [hep-ph/9609276].

\bibitem{Bilenky:2001rz} 
  S.~M.~Bilenky, S.~Pascoli and S.~T.~Petcov,
Phys.\ Rev.\ D {\bf 64}, 053010 (2001) 
[hep-ph/0102265].

\bibitem{Czakon:2001uh} 
M.~Czakon, J.~Gluza, J.~Studnik and M.~Zralek,
Phys.\ Rev.\ D {\bf 65}, 053008 (2002)  [hep-ph/0110166].

\bibitem{Pascoli:2001by} 
S.~Pascoli, S.~T.~Petcov and L.~Wolfenstein,
Phys.\ Lett.\ B {\bf 524}, 319 (2002)  [hep-ph/0110287].

\bibitem{Barger:2002vy} 
V.~Barger, S.~L.~Glashow, P.~Langacker and D.~Marfatia,
Phys.\ Lett.\ B {\bf 540}, 247 (2002)  [hep-ph/0205290].

\bibitem{Nunokawa:2002iv} 
  H.~Nunokawa, W.~J.~C.~Teves and R.~Zukanovich Funchal,
Phys.\ Rev.\ D {\bf 66}, 093010 (2002)
[hep-ph/0206137].

\bibitem{Pascoli:2002qm} 
  S.~Pascoli, S.~T.~Petcov and W.~Rodejohann,
Phys.\ Lett.\ B {\bf 549}, 177 (2002)
[hep-ph/0209059].

\bibitem{Deppisch:2004kn} 
  F.~Deppisch, H.~Pas and J.~Suhonen,
Phys.\ Rev.\ D {\bf 72}, 033012 (2005)
  [hep-ph/0409306].

\bibitem{Joniec:2004mx} 
  A.~Joniec and M.~Zralek,
Phys.\ Rev.\ D {\bf 73}, 033001 (2006)
[hep-ph/0411070].


\bibitem{Pascoli:2005zb} 
  S.~Pascoli, S.~T.~Petcov and T.~Schwetz,
Nucl.\ Phys.\ B {\bf 734}, 24 (2006)
[hep-ph/0505226].

\bibitem{Choubey:2005rq} 
  S.~Choubey and W.~Rodejohann,
Phys.\ Rev.\ D {\bf 72}, 033016 (2005)
[hep-ph/0506102].

\bibitem{Simkovic:2012hq} 
 F.~Simkovic, S.~M.~Bilenky, A.~Faessler and T.~Gutsche,
 Phys.\ Rev.\ D {\bf 87}, 073002 (2013)
 [arXiv:1210.1306 [hep-ph]].

\bibitem{Ade:2013zuv}
P.~A.~R.~Ade {\it et al.}  [Planck Collaboration],
Astron.\ Astrophys.\  {\bf 571}, A16 (2014)
[arXiv:1303.5076 [astro-ph.CO]].

\bibitem{Abazajian:2011dt}
  K.~N.~Abazajian, E.~Calabrese, A.~Cooray, F.~De Bernardis, S.~Dodelson, A.~Friedland, G.~M.~Fuller and S.~Hannestad {\it et al.},
Astropart.\ Phys.\  {\bf 35}, 177 (2011)
[arXiv:1103.5083 [astro-ph.CO]].


\bibitem{Battye:2013xqa} 
  R.~A.~Battye and A.~Moss,
  Phys.\ Rev.\ Lett.\  {\bf 112}, 051303 (2014)
  [arXiv:1308.5870 [astro-ph.CO]].


\bibitem{Agostini:2013mzu}
  M.~Agostini {\it et al.}  [GERDA Collaboration], 
Phys.\ Rev.\ Lett.\  {\bf 111}, 122503 (2013) 
[arXiv:1307.4720 [nucl-ex]].

\bibitem{Gando:2012zm}
A.~Gando {\it et al.}  [KamLAND-Zen Collaboration],
Phys.\ Rev.\ Lett.\  {\bf 110}, 062502 (2013)
[arXiv:1211.3863 [hep-ex]].

\bibitem{Albert:2014awa}
  J.~B.~Albert {\it et al.}  [EXO-200 Collaboration],
  Nature {\bf 510} 229, (2014)
  [arXiv:1402.6956 [nucl-ex]].

\bibitem{Auger:2012ar}
  M.~Auger {\it et al.}  [EXO Collaboration],
Phys.\ Rev.\ Lett.\  {\bf 109}, 032505 (2012)
[arXiv:1205.5608 [hep-ex]].

\bibitem{Arnaboldi:2002du}
C.~Arnaboldi {\it et al.}  [CUORE Collaboration],
{\it  Nucl.\ Instrum.\ Meth.}\ A {\bf 518}, 775 (2004)
[hep-ex/0212053].

\bibitem{Hartnell:2012qd}
  J.~Hartnell [SNO+ Collaboration],
J.\ Phys.\ Conf.\ Ser.\  {\bf 375}, 042015 (2012)
[arXiv:1201.6169 [physics.ins-det]].


\bibitem{Gaitskell:2003zr}
R.~Gaitskell {\it et al.}  [Majorana Collaboration],
  nucl-ex/0311013.

\bibitem{Arnold:2010tu}
  R.~Arnold {\it et al.}  [SuperNEMO Collaboration],
Eur.\ Phys.\ J.\ C {\bf 70}, 927 (2010)
 [arXiv:1005.1241 [hep-ex]].

\bibitem{Alvarez:2011my}
  V.~Alvarez {\it et al.}  [NEXT Collaboration],
  arXiv:1106.3630 [physics.ins-det].

\bibitem{GomezCadenas:2011it} 
  J.~J.~Gomez-Cadenas, J.~Martin-Albo, M.~Mezzetto, F.~Monrabal and M.~Sorel,
Riv.\ Nuovo Cim.\  {\bf 35}, 29 (2012)
[arXiv:1109.5515 [hep-ex]].

\bibitem{Giuliani:2012zu} 
  A.~Giuliani and A.~Poves,
Adv.\ High Energy Phys.\  {\bf 2012}, 857016 (2012).

\bibitem{Cremonesi:2013bma} 
  O.~Cremonesi and M.~Pavan,
arXiv:1310.4692 [physics.ins-det].

\bibitem{Rodin:2003eb}
  V.~A.~Rodin, A.~Faessler, F.~Simkovic and P.~Vogel,
Phys.\ Rev.\ C {\bf 68},044302 (2003) 
[nucl-th/0305005].

\bibitem{Rodin:2007fz}
  V.~A.~Rodin, A.~Faessler, F.~Simkovic and P.~Vogel,
Nucl.\ Phys.\ A {\bf 766}, 107 (2006)
[Erratum-ibid.\ A {\bf 793}, 213 (2007)]
[arXiv:0706.4304 [nucl-th]].

\bibitem{Caurier:1996zz}
  E.~Caurier, F.~Nowacki, A.~Poves and J.~Retamosa,
Phys.\ Rev.\ Lett.\  {\bf 77}, 1954 (1996).

\bibitem{Caurier:2007wq} 
  E.~Caurier, J.~Menendez, F.~Nowacki and A.~Poves,
 Phys.\ Rev.\ Lett.\  {\bf 100}, 052503 (2008)
 [arXiv:0709.2137 [nucl-th]].

\bibitem{Simkovic:2009pp}
  F.~Simkovic, A.~Faessler, H.~Muther, V.~Rodin and M.~Stauf,
Phys.\ Rev.\ C {\bf 79}, 055501 (2009)
[arXiv:0902.0331 [nucl-th]].

\bibitem{Kortelainen:2007rh} 
  M.~Kortelainen and J.~Suhonen,
Phys.\ Rev.\ C {\bf 75}, 051303 (2007)
[arXiv:0705.0469 [nucl-th]].

\bibitem{Kortelainen:2007mn} 
  M.~Kortelainen and J.~Suhonen,
Phys.\ Rev.\ C {\bf 76}, 024315 (2007)
 [arXiv:0708.0115 [nucl-th]].

\bibitem{Barea:2009zza} 
  J.~Barea and F.~Iachello,
Phys.\ Rev.\ C {\bf 79}, 044301 (2009).

\bibitem{Barea:2012zz} 
  J.~Barea, J.~Kotila and F.~Iachello,
Phys.\ Rev.\ Lett.\  {\bf 109}, 042501 (2012).

\bibitem{Barea:2013bz} 
  J.~Barea, J.~Kotila and F.~Iachello,
Phys.\ Rev.\ C {\bf 87}, 014315 (2013)
  [arXiv:1301.4203 [nucl-th]].

\bibitem{Menendez:2008jp} 
  J.~Menendez, A.~Poves, E.~Caurier and F.~Nowacki,
 Nucl.\ Phys.\ A {\bf 818}, 139 (2009)
  [arXiv:0801.3760 [nucl-th]].


\bibitem{Gando:2013nba} 
A.~Gando {\it et al.}  [KamLAND Collaboration],
Phys.\ Rev.\ D {\bf 88}, 033001 (2013)
[arXiv:1303.4667 [hep-ex]].




\bibitem{Kettell:2013eos}
  S.~Kettell, J.~Ling, X.~Qian, M.~Yeh, C.~Zhang, C.~-J.~Lin, K.~-B.~Luk and R.~Johnson {\it et al.},
arXiv:1307.7419 [hep-ex].


\bibitem{Seo:2013swa}
S.~-H.~Seo,
Nucl.\ Phys.\ Proc.\ Suppl.\  {\bf 237},65 (2013) 
See also 
http://home.kias.re.kr/MKG/h/reno50.

\bibitem{T2K} 
K.~Abe {\it et al.}  [T2K Collaboration],
  Phys.\ Rev.\ Lett.\  {\bf 107}, 041801 (2011)
  [arXiv:1106.2822 [hep-ex]].

\bibitem{An:2012bu}
  F.~P.~An {\it et al.}  [Daya Bay Collaboration],
 Chin.\  Phys.\ C {\bf 37}, 011001 (2013)
  [arXiv:1210.6327 [hep-ex]].

\bibitem{RENO}
H.~Seo, 
{\it Recent Results from RENO}, 
Talk at XXIV Workshop on Weak Interactions and Neutrinos (WIN 2013), 
September 16-21, Natal, Brazil.


\bibitem{Wolf:2008hf}
  J.~Wolf [KATRIN Collaboration],
Nucl.\ Instrum.\ Meth.\ A {\bf 623}, 442 (2010)
[arXiv:0810.3281 [physics.ins-det]].

\bibitem{KATRIN}
http://www-ik.fzk.de/~katrin/publications/documents/DesignReport2004-12Jan2005.pdf

\bibitem{Machado:2013kya} 
  P.~A.~N.~Machado, H.~Minakata, H.~Nunokawa and R.~Zukanovich Funchal,
  JHEP {\bf 1405}, 109 (2014)
  [arXiv:1307.3248].

\bibitem{Winter:2003ye} 
  W.~Winter,
 Phys.\ Rev.\ D {\bf 70}, 033006 (2004)
[hep-ph/0310307].

\bibitem{Huber:2004gg} 
  P.~Huber, M.~Lindner and W.~Winter,
JHEP {\bf 0505}, 020 (2005)
[hep-ph/0412199].

\bibitem{Fogli:2004as}
  G.~L.~Fogli, E.~Lisi, A.~Marrone, A.~Melchiorri, A.~Palazzo, P.~Serra and J.~Silk,
  Phys.\ Rev.\ D {\bf 70}, 113003 (2004)
  [hep-ph/0408045].


\bibitem{Agashe:2014kda} 
K.~A.~Olive {\it et al.}  [Particle Data Group Collaboration],
Chin.\ Phys.\ C {\bf 38}, 090001 (2014).




\bibitem{Pantis:1996py} 
  G.~Pantis, F.~Simkovic, J.~D.~Vergados and A.~Faessler,
  Phys.\ Rev.\ C {\bf 53}, 695 (1996)
  [nucl-th/9612036].

\bibitem{Kotila:2012zza} 
  J.~Kotila and F.~Iachello,
  Phys.\ Rev.\ C {\bf 85}, 034316 (2012)
  [arXiv:1209.5722 [nucl-th]].

\bibitem{Riemer-Sorensen:2013jsa} 
  S.~Riemer-Sorensen, D.~Parkinson and T.~M.~Davis,
  arXiv:1306.4153 [astro-ph.CO].

\bibitem{Parkinson:2012vd}
  D.~Parkinson, S.~Riemer-Sorensen, C.~Blake, G.~B.~Poole, T.~M.~Davis, S.~Brough, M.~Colless and C.~Contreras {\it et al.},
  Phys.\ Rev.\ D {\bf 86} (2012) 103518
  [arXiv:1210.2130 [astro-ph.CO]].


\bibitem{Wyman:2013lza}
 M.~Wyman, D.~H.~Rudd, R.~A.~Vanderveld and W.~Hu,
  Phys.\ Rev.\ Lett.\  {\bf 112}, 051302 (2014)
  [arXiv:1307.7715 [astro-ph.CO]].

\bibitem{Hamann:2013iba} 
  J.~Hamann and J.~Hasenkamp,
  JCAP {\bf 1310}, 044 (2013)
  [arXiv:1308.3255 [astro-ph.CO]].

\bibitem{Carbone:2010ik}
  C.~Carbone, L.~Verde, Y.~Wang and A.~Cimatti,
 JCAP {\bf 1103}, 030 (2011)
[arXiv:1012.2868 [astro-ph.CO]].

\bibitem{Kitching:2008dp}
  T.~D.~Kitching, A.~F.~Heavens, L.~Verde, P.~Serra and A.~Melchiorri,
Phys.\ Rev.\ D {\bf 77}, 103008 (2008)
[arXiv:0801.4565 [astro-ph]].

\bibitem{Amendola:2012ys} 
  L.~Amendola {\it et al.}  [Euclid Theory Working Group Collaboration],
Living Rev.\ Rel.\  {\bf 16}, 6 (2013)
[arXiv:1206.1225 [astro-ph.CO]].

\bibitem{Hamann:2012fe} 
  J.~Hamann, S.~Hannestad and Y.~Y.~Y.~Wong,
JCAP {\bf 1211}, 052 (2012)
[arXiv:1209.1043 [astro-ph.CO]].

\bibitem{Leistedt:2014sia} 
  B.~Leistedt, H.~V.~Peiris and L.~Verde,
  Phys.\ Rev.\ Lett.\  {\bf 113}, 041301 (2014)
  [arXiv:1404.5950 [astro-ph.CO]].

\bibitem{Otten:2008zz}
  E.~W.~Otten and C.~Weinheimer,
  Rept.\ Prog.\ Phys.\  {\bf 71} (2008) 086201
  [arXiv:0909.2104 [hep-ex]].

\bibitem{Kraus:2004zw}
C.~Kraus, B.~Bornschein, L.~Bornschein, J.~Bonn, B.~Flatt, 
A.~Kovalik, B.~Ostrick and E.~W.~Otten {\it et al.},
Eur.\ Phys.\ J.\ C {\bf 40}, 447 (2005) [hep-ex/0412056].

\bibitem{Aseev:2011dq}
V.~N.~Aseev {\it et al.}  [Troitsk Collaboration],
Phys.\ Rev.\ D {\bf 84}, 112003 (2011)
[arXiv:1108.5034 [hep-ex]].

\bibitem{Pascoli:2002xq} 
  S.~Pascoli and S.~T.~Petcov,
  Phys.\ Lett.\ B {\bf 544}, 239 (2002)
  [hep-ph/0205022].
 
\bibitem{Pascoli:2002ae} 
  S.~Pascoli, S.~T.~Petcov and W.~Rodejohann,
  Phys.\ Lett.\ B {\bf 558}, 141 (2003)
  [hep-ph/0212113].

\bibitem{Petcov:2004wz} 
  S.~T.~Petcov,
  New J.\ Phys.\  {\bf 6}, 109 (2004).

\bibitem{Lindner:2005kr} 
  M.~Lindner, A.~Merle and W.~Rodejohann,
  Phys.\ Rev.\ D {\bf 73}, 053005 (2006)
  [hep-ph/0512143].

\bibitem{Vissani:1999tu}
  F.~Vissani,
  JHEP {\bf 9906}, 022 (1999)   [hep-ph/9906525].

\bibitem{Capozzi:2013csa} 
F.~Capozzi, G.~L.~Fogli, E.~Lisi, A.~Marrone, D.~Montanino and A.~Palazzo,
arXiv:1312.2878 [hep-ph].

\bibitem{Elliott:2002xe} 
  S.~R.~Elliott and P.~Vogel,
Ann.\ Rev.\ Nucl.\ Part.\ Sci.\  {\bf 52}, 115 (2002)
  [hep-ph/0202264].

\bibitem{Host:2007wh} 
  O.~Host, O.~Lahav, F.~B.~Abdalla and K.~Eitel,
 Phys.\ Rev.\ D {\bf 76}, 113005 (2007)
  [arXiv:0709.1317 [hep-ph]].

\bibitem{Hannestad:2007tu} 
  S.~Hannestad,
  arXiv:0710.1952 [hep-ph].

\bibitem{Bahcall:2003ce} 
  J.~N.~Bahcall and C.~Pena-Garay,
  JHEP {\bf 0311}, 004 (2003)
  [hep-ph/0305159].


\bibitem{Minakata:2004jt} 
  H.~Minakata, H.~Nunokawa, W.~J.~C.~Teves and R.~Zukanovich Funchal,
Phys.\ Rev.\ D {\bf 71}, 013005 (2005)
[hep-ph/0407326].

\bibitem{Bandyopadhyay:2004cp} 
 A.~Bandyopadhyay, S.~Choubey, S.~Goswami and S.~T.~Petcov,
Phys.\ Rev.\ D {\bf 72}, 033013 (2005)  [hep-ph/0410283].

\bibitem{Ge:2012wj} 
  S.~-F.~Ge, K.~Hagiwara, N.~Okamura and Y.~Takaesu,
JHEP {\bf 1305}, 131 (2013)
  [arXiv:1210.8141 [hep-ph]].

\bibitem{Capozzi:2013psa} 
 F.~Capozzi, E.~Lisi and A.~Marrone,
  Phys.\ Rev.\ D {\bf 89}, 013001 (2014)
  [arXiv:1309.1638 [hep-ph]].

\bibitem{Minakata:2014jba}
  H.~Minakata, H.~Nunokawa and A.~A.~Quiroga,
  arXiv:1402.6014v1 [hep-ph].

\bibitem{Avignone:2005cs} 
F.~T.~Avignone, G.~S.~King and Y.~.G.~Zdesenko,
New J.\ Phys.\  {\bf 7}, 6 (2005).

\bibitem{Avignone:2007fu} 
  F.~T.~Avignone, III, S.~R.~Elliott and J.~Engel,
Rev.\ Mod.\ Phys.\  {\bf 80}, 481 (2008)
[arXiv:0708.1033 [nucl-ex]].

\bibitem{Moe:1991ku} 
  M.~K.~Moe,
Nucl.\ Phys.\ Proc.\ Suppl.\  {\bf 19}, 158 (1991).

\bibitem{Dodelson:2014tga} 
S.~Dodelson and J.~Lykken,
arXiv:1403.5173 [astro-ph.CO].
 
\end{thebibliography}
\end{document}